\documentclass[aip,onecolumn,amsmath,amssymb,pof,preprint,floatfix]{revtex4-1}
\usepackage{graphicx}
\usepackage{graphics}
\usepackage{color}
\usepackage{rotating}

\def \i{\`{\i}~}

\def\ld{$-$}

\begin{document}

\title{
Intrinsic filtering errors of Lagrangian particle tracking\\
in LES flow fields
}

\author{F. Bianco$^{1,2}$, S. Chibbaro$^2$, C. Marchioli$^{1,3}$, M.V. Salvetti$^4$, A. Soldati$^{1,3}$}
\affiliation{$^1$Energy Technology Dept.,
University of Udine, 33100 Udine -- Italy\\
$^2$Institut Jean Le Rond D'Alembert UMR 7190 CNRS, 
University Pierre et Marie Curie,  
75252 Paris --  France\\
$^3$ Dept. Fluid Mechanics, CISM
(Centro Interdipartimentale di Fluidodinamica e Idraulica),
33100 Udine -- Italy\\
$^4$Aerospace Engineering Dept., University of Pisa, 56100 Pisa -- Italy 
}%
\date{\today}
%


\begin{abstract}
Large-Eddy Simulations (LES) of
two-phase turbulent flows exhibit quantitative differences
in particle statistics if compared to Direct Numerical
Simulations (DNS) which, in the context of the present study,
is considered the exact reference case.
Differences are primarily due to filtering, a fundamental intrinsic
feature of LES. Filtering the fluid velocity field yields
approximate computation of the forces
acting on particles and, in turn, trajectories that are
inaccurate when compared to those of DNS.
In this paper, we focus precisely on the filtering error for
which we quantify a lower bound. To this aim, we
use a DNS database
of inertial particle dispersion in turbulent channel flow
and we perform a-priori tests in which the error purely due
to filtering is singled out removing error accumulation
effects, which would otherwise lead to progressive divergence
between DNS and LES particle trajectories.
By applying filters of different
type and width at varying particle inertia, we characterize the statistical properties
of the filtering error as a function of the wall distance.
Results show that filtering error is stochastic and has
a non-Gaussian distribution.
In addition, the distribution of the filtering error depends strongly
on the wall-normal coordinate being maximum in the buffer
region.
Our findings provide insight on the effect of subgrid-scale velocity
field on the force driving the particles, and establish the requirements
which a LES model must satisfy to predict correctly the velocity and
the trajectory of inertial particles.
\end{abstract}
%
\maketitle
%
\section{Introduction}
\label{intro}
The dispersion of small inertial particles in inhomogeneous turbulent
flow
has been long recognized as crucial in a number of industrial
applications and environmental phenomena. Examples include mixing,
combustion, depulveration, spray dynamics, pollutant dispersion
or cloud dynamics.
In all these problems accurate predictions
are important, yet not trivial to obtain because
of the complex phenomenology that controls turbulent particle dynamics.

Direct Numerical Simulations (DNS)
of turbulence coupled with Lagrangian Particle Tracking (LPT)
\cite{hit2,ms02,hit1,fs06,re01,Soldati2009,mgss03,uo96}
have demonstrated their capability to capture the
physics of particle dynamics in relation with turbulence dynamics
and have highlighted the key role played by inertial
clustering and preferential concentration
in determining the rates of particle interaction, settling,
deposition and entrainment.
Due to the computational requirements of DNS,
however, analysis of applied problems characterized by
complex geometries
and high Reynolds numbers demands alternative approaches,
among which Large-Eddy Simulation
(LES) is gaining in popularity, especially for
cases where the large flow scales control particle motion
(e.g. \cite{apte,okong}).

LES is based on a filtering approach in which the unresolved
Sub-Grid Scales (SGS) of turbulence are modeled. 
A major issue associated with LES of
particle-laden flows is modeling the effect of these
scales on particle dynamics.
Flow anisotropy adds a further complicacy.
In particular, recent studies
on particle dispersion in wall-bounded
turbulence
\cite{kv05,k06,Marchioli2008,Khan2010} have demonstrated that
use of the filtered fluid velocity in the equation of particle
motion with no model for
the effect of the SGS
fluid velocity fluctuations
leads to significant underestimation of
preferential concentration and, consequently, to weaker
deposition fluxes and lower near-wall accumulation.
Even though several studies have demonstrated that
predictions may be improved using techniques
like filter inversion or approximate deconvolution
\cite{kv05,k06,sm05,Gobert2010,Marchioli2008b},
the amount of velocity fluctuations that can be reintroduced
in the particle equations does not
ensure a quantitative {\em replica} of DNS results.
A further effort is required to model the scales smaller
than the filter width but extremely significant for
particle dynamics, which are inevitably lost in the
filtering procedure \cite{Marchioli2008b}.
Attempts to model
SGS effects on particles have been made
using fractal interpolation \cite{Marchioli2008b} or
kinematic simulations \cite{Khan2010} in channel flow.
Also these approaches, however,
do not ensure complete removal of inaccuracies
in the quantitative prediction of local particle
segregation and accumulation (particularly in the near-wall
region).
Stochastic models have also
been proposed, yet we are only aware of validations in
homogeneous isotropic turbulence \cite{sm06,Pozorski2009,Gobert2010}:
these
models have still to be assessed in
non-homogeneous anisotropic flows.

To devise accurate and reliable SGS models for LPT,
precise quantification of filtering effects
on particle dynamics is clearly
a necessary step. Aim of this paper is to propose a simple
procedure to characterize the subgrid
error \cite{kv05} purely due to filtering of the fluid
velocity field.
This error, referred to as pure filtering error hereinafter,
can not be avoided
but potentially can be corrected.
We thus focus on an ideal situation in which
LES provides the exact dynamics of the resolved velocity
field (\textit{ideal LES}). In ideal LES, 
modeling and numerical errors (incurred because a \textit{real LES} provides
only an approximation of the filtered velocity)
and interpolation errors (on
coarse-grained domains, for instance) \cite{kv05} are
assumed negligible, and a lower bound for the pure
filtering error can be identified.
Peculiar feature of the procedure proposed here is that it
removes time
accumulation of pure filtering errors on particle trajectories.
Error accumulation originates from inaccurate estimation of
the forces acting on particles obtained when the filtered
fluid velocity is supplied to the equation of particle motion.
As a result, particle
trajectories in LES fields progressively diverge from
particle trajectories in DNS fields, considered
as the \textit{exact} reference for the present study:
the flow
fields {\em seen} by the particles become less and less
correlated and the forces acting on particles are evaluated
at increasingly different locations.
We remark that the effects due to these two errors can not
be singled out easily when comparing statistics of particle velocity
and concentration obtained from LES to the reference DNS data.
To remove this effect, we perform here a-priori tests in which:
{\em (i)} particle trajectories are computed from the DNS fields,
{\em (ii)} the DNS fields are coarse-grained through filtering and, 
{\em (iii)} particles are forced to evolve in the filtered DNS fields
along the DNS trajectories.
%
A similar a-priori analysis has been performed recently
to evaluate pure filtering error and time accumulation error
for the case of tracer particle dispersion in
homogeneous isotropic turbulence \cite{Calzavarini2010}.
In this paper, we focus on inertial particles in turbulent
channel flow.

A-priori evaluation of the statistical properties of the
filtering error provides useful information
about the key features that should be incorporated in SGS
models for LPT to compensate for such error.
Statistics are computed applying filters
of different type and different widths, corresponding roughly
to varying amounts of resolved flow energy, and considering
point particles with different inertia that obey one-way
coupled Stokes dynamics. One point at issue is indeed whether
SGS models for LPT should take into account inertial effects
explicitly.
Another point to be explored is whether the characteristics
of SGS models for LPT should adapt ``dynamically'' to respond
to the local inhomogeneity and anisotropy of the flow.
Our study aims at addressing these issues through statistical
characterization of the effects of the unresolved velocity field
on the force driving the particles.
Possible ways to improve prediction of LES applied to
turbulent dispersed flow should be benchmarked against
this type of statistical information at the sub-grid level.

The paper is organized as follows. The physical problem and
the numerical methodology are described in Sec. \ref{meth}.
The statistical moments and the probability distribution functions
of the filtering error are presented in Sec. \ref{statistics} and
in Sec. \ref{PDF}, respectively.
Finally, concluding remarks are given in Sec. \ref{conclusions}.

\section{Physical Problem and Numerical Methodology}
\label{meth}

\subsection{Particle-laden turbulent channel flow}

The reference flow configuration consists of two infinite
flat parallel walls: the origin of the coordinate system is located
at the center of the channel and the $x-$, $y-$ and $z-$ axes point
in the streamwise, spanwise and wall-normal directions respectively.
Periodic boundary conditions are imposed on the fluid velocity field
in $x$ and $y$, no-slip boundary conditions
are imposed at the walls. In the present study, we consider air (assumed
to be incompressible and Newtonian)
with density $\rho = 1.3~kg~m^{-3}$ and kinematic
viscosity $\nu = 15.7{\times}10^{-6}~m^{2}~s^{-1}$.
The flow is driven by a mean pressure gradient, 
and the shear Reynolds number is
$Re_{\tau} = u_{\tau}h/\nu=150$, based on the shear (or friction) velocity, $u_{\tau}$,
and on the half channel height, $h$.
The shear velocity is defined as $u_{\tau} = (\tau_{w}/\rho)^{1/2}$,
where $\tau_{w}$ is the mean shear stress at the wall.

Particles are modeled as pointwise, non-rotating rigid spheres
with density $\rho_p=10^3~kg~m^{-3}$, and are injected
into the flow at concentration low enough to
consider dilute system conditions (no inter-particle
collisions) and one-way coupling between the two phases
(no turbulence modulation by particles).
The motion of
particles is described by a set of ordinary differential
equations for particle velocity and position.
For particles much heavier than the fluid ($\rho_{p}/\rho \gg 1$)
Elghobashi and Truesdell \cite{et92} have shown that the most
significant forces are Stokes drag and buoyancy
and that Basset force can be neglected being
an order of magnitude smaller.
In the present simulations, the aim is to minimize the
number of degrees of freedom by keeping the simulation setting
as simplified as possible; thus
the effect of gravity has also been
neglected. 
With these assumptions, a
simplified version of the Basset-Bousinnesq-Oseen
equation \cite{Gatignol} is obtained.
In vector form:
\begin{eqnarray}
\label{partpos}
\frac{d{\bf x}_p}{dt} = {\bf v}_p~,
\end{eqnarray}
\vspace{-0.6cm}
\begin{eqnarray}
\label{partvel}
\frac{d{\bf v}_p}{dt} =
\frac{\mathbf{u}_s -\mathbf v_p }{\tau_p} (1 + 0.15 Re_p^{0.687})~.
\end{eqnarray}
In Eqns. (\ref{partpos}) and (\ref{partvel}), ${\bf x}_p$ is particle position,
${\bf v}_p$ is particle velocity, 
${\bf u}_s={\bf u}({\bf x}_p(t),t)$ is the fluid velocity
at the particle position.
$\tau_p = \rho_{p}d^{2}_{p}/18\mu$ is the particle relaxation time, and
$Re_p = d_{p}|{\bf v}_p - {\bf u}_s| \rho /\mu$ is the particle Reynolds
number, $d_{p}$ and $\mu$ being the particle diameter and the fluid
dynamic viscosity respectively.

\subsection{DNS/LPT methodology}

The Eulerian flow field is obtained using a pseudo-spectral DNS flow solver
which discretizes the governing equations (Continuity and
Navier-Stokes for incompressible flow) by transforming the field variables
into wavenumber space, using Fourier representations for the periodic
streamwise and spanwise directions and a Chebyshev representation for the
wall-normal (non-homogeneous) direction.
A two level, explicit
Adams-Bashforth scheme for the non-linear terms, and an implicit
Crank-Nicolson method for the viscous terms are employed
for time advancement.
Further details of the method can be found in previous articles
(e.g. Pan and Banerjee\cite{PB_POF_96}).
Calculations were performed
on a computational domain of size
$L_x \times L_y \times L_z = 4 \pi h \times 2 \pi h \times 2 h$
in $x$, $y$ and $z$ respectively, corresponding to
$1885\times942\times300$ in wall units. Wall
units are obtained combining $u_{\tau}$, $\nu$ and $\rho$.
The computational domain is discretized in physical space
with $N_x \times N_y \times N_z = 128\times128\times129$
grid points (corresponding to $128\times128$
Fourier modes and to 129 Chebyshev coefficients in the wavenumber space). 
This is the minimum number of grid points required in each direction
to ensure that the grid spacing is always smaller than the smallest
flow scale and that the limitations imposed
by the point-particle approach are satisfied. \cite{Marchioli2008}

To calculate particle trajectories a Lagrangian tracking routine is
coupled to the flow solver. The routine solves for Eqns. (\ref{partpos})
and (\ref{partvel}) using a $4^{th}$-order Runge-Kutta scheme for time
advancement and $6^{th}$-order Lagrangian polynomials for fluid velocity
interpolation at the particle location.
At the beginning of the simulation, particles
are distributed randomly within
the computational domain and their initial
velocity is set equal to that of the fluid
at the particle initial position.
Periodic boundary conditions are imposed on particles
moving outside the computational domain in the
homogeneous directions,
perfectly-elastic collisions at the smooth walls are assumed
when the particle center is at a distance lower than one
particle radius from the wall.
For the simulations presented here,
large samples of $10^{5}$ particles, characterized by
different response times, are considered. When
the particle response time is made
dimensionless using wall variables,
the Stokes number for each particle
set is obtained as $St=\tau_p/\tau_f$
where $\tau_f=\nu/u_{\tau}^2$ is the
viscous timescale of the flow.
Three different sets of particles corresponding to values of the Stokes number
$St=1$, $5$ and $25$ have been considered in this study:
Table~\ref{table:part} shows the relevant physical parameters
of each set.
The total tracking time in dimensionless wall units (identified
hereinafter by superscript +) is $\Delta T^+ = 21150$
for all particle sets \cite{Marchioli2008},
the timestep size for particle tracking
being equal to the timestep size used for the fluid: $\delta t^+ = 0.045$.
\subsection{A-priori LES methodology and computation of the filtering error}
\label{IC}
In the a-priori tests, LPT is carried out replacing ${\bf u}_s$ in Eq.
(\ref{partvel}) with the filtered fluid
velocity field, $\bar{{\bf u}}({\bf x_p},t)$. This field is
obtained through explicit filtering of the DNS velocity by
either a cut-off filter or a top-hat filter. Both filters are applied
in the wave number space to velocity components in the homogeneous
streamwise and spanwise directions:
{\small \begin{eqnarray}
\label{cut-off}
\bar{u}_i ({\bf x},t) = FT^{-1} \left\{
\begin{array}{ll}
{\hat G}(\kappa_1) \cdot {\hat G}(\kappa_2) \cdot
\hat{u}_i (\kappa_1,\kappa_2,z,t)~~{{\text{if}}~| \kappa_j | \le |\kappa_c|
~{\text{with}}~j=1,2}\\
0 \hspace{5.cm} \text{otherwise}
\end{array}
\right.
\end{eqnarray}}
where $FT$ is the 2D Fourier Transform, $\kappa_c= \pi / \Delta$ is the
cutoff wave number ($\Delta$ being the filter width in the physical
space), $\hat{u}_i (\kappa_1,\kappa_2,z,t)$
is the Fourier transform of the DNS fluid velocity field, namely
$\hat{u}_i (\kappa_1,\kappa_2,z,t)=FT [{u}_i ({\bf x},t) ]$ and
${\hat G}(\kappa_j)$ is
the filter transfer function:
\begin{eqnarray}
\label{transfer-function}
G(\kappa_j) = \left\{
\begin{array}{ll}
1~~~~~~~~~~~~~~~~~~~{\text{for the cut-off filter}}\\
\frac{sin(\kappa_j \Delta/2)}{\kappa_j \Delta/2} ~~~~~~~~~\text{for the top-hat filter}
\end{array}
\right.
\end{eqnarray}
Three different filter widths are considered,
which provide a grid Coarsening Factor (CF) in each homogeneous
direction of 2, 4 and 8 with respect to DNS, corresponding to $64 \times 64$,
$32 \times 32$ and $16 \times 16$ Fourier modes respectively.
Note that CF=2 and CF=4 yield grid resolutions that are
commonly used in LES, whereas CF=8 corresponds to a very coarse
resolution characteristic of under-resolved LES.
Data are not filtered in the wall-normal direction,
since the wall-normal resolution in LES is often DNS-like. \cite{pope-faq}

The pure filtering error is computed under the ideal assumption that all further
sources of error affecting particle tracking in LES can be disregarded
and that the exact dynamics of the resolved velocity field are
available. Errors due to the SGS model for the fluid and to numerics are thus neglected and a lower bound for the filtering error is identified
by removing error accumulation effects due to progressive divergence
between DNS and LES particle trajectories.
Time accumulation of the pure filtering error is prevented by computing particle
trajectories in DNS fields and then forcing particles to evolve in
filtered DNS fields along the DNS trajectories.
At each time step $n$ and for each particle $k$ we thus impose:
\begin{eqnarray}
\label{cond_corr}
\mathbf{x}_{p,k}^{LES}(t^n)=\mathbf{x}_{p,k}^{DNS}(t^n) \equiv \mathbf{x}_{p,k}(t^n) \; \; ; \; \;
\mathbf{v}_{p,k}^{LES}(t^n)=\mathbf{v}_{p,k}^{DNS}(t^n) \equiv \mathbf{v}_{p,k}(t^n)~~,
\end{eqnarray}
where superscripts $LES$ and $DNS$ identify particle position and velocity
in a-priori LES and in DNS, respectively.
The pure filtering error made on the $k$-th particle
at time $t^n$ is computed as:
\begin{eqnarray}
\label{corr_term}
\mathbf{\delta u} \equiv
\mathbf{\delta u}(\mathbf{x}_{p,k}(t^n),t^n) =
\mathbf{u}(\mathbf{x}_{p,k}(t^n),t^n) - \bar{\mathbf{u}}(\mathbf{x}_{p,k}(t^n),t^n)
\equiv \mathbf{u}_s - \bar{\mathbf{u}}_s~~.
\end{eqnarray}
Due to the wall-normal dependence of the fluid velocity
(and of $\mathbf{u}_s$ in particular),
it is not straightforward to select a characteristic velocity to normalize
$\delta {\bf u}$. We thus decided to provide measures of the
absolute filtering error rather than the percent filtering error,
using Eq. (\ref{corr_term}) to quantify exactly the filtering error affecting
the fluid velocity seen by the particles.
This quantification relies on an idealized situation which serves the purpose of isolating the contribution due to filtering
of the velocity field to the total error associated with LPT in LES.
Note also that $\mathbf{\delta u}$ is computed along the ``exact'' trajectory
of each particle, that is at ${\bf x}_{p,k}^{DNS}(t^n)$. As shown
in the following, the corresponding behavior observed for $\mathbf{\delta u}$
may be significantly different from the Eulerian measure of the filtering
error usually found in the literature, which is obtained as difference
between DNS and LES velocities at fixed points in space.

Computation and statistical characterization of $\mathbf{\delta u}$ were
carried out for all particle sets reported in Table \ref{table:part} and
applying filters of
different type and width. The averaging time window required for converged
statistics was $\Delta t^+ \simeq 8285$.
\section{Statistical moments of the pure filtering error}
\label{statistics}

The Eulerian statistical moments of the filtering error, quantified
here by its lower bound $\mathbf{\delta u}$ (Eq. (\ref{corr_term})), are
presented and analyzed in this section.
Statistics were computed by dividing the computational domain in
wall-parallel slabs having dimensions $L_x$, $L_y$ and $\Delta_z(i)$ with
$i=1, N_z$, where $\Delta_z(i)$ is the difference between two adjacent
Chebyshev collocation points.
Ensemble averaging is carried out over particles that, at a given
time instant, are located inside the same slab; time averaging is
performed over the time interval $\Delta t^+$.
All quantities shown in this section are non dimensional and expressed
in wall units.
\subsection{Influence of particle inertia on the filtering error}
\label{statistics_inertia}
To analyze the effect of particle inertia on the behavior of
$\delta {\bf u}$, we compare statistics relative to all particle sets
but we focus one filter type (cut-off) and one filter width (CF=4).
This filter width is chosen as it removes a significant amount
of spatial information on the turbulent structures which interact
with particles without generating overly under-resolved fields.
The effects due to different filter types and widths will be investigated
in Sec. \ref{statistics_filter}.

Figure \ref{mean_IC_inertia} shows the mean absolute
value $\langle \delta {\bf u} \rangle$ of the filtering error
along the streamwise, spanwise and normal directions
as a function of the distance from the wall.
Hereinafter $\langle \cdot \rangle$ identifies time- and ensemble-averaged quantities.
The qualitative behavior of the streamwise component $\langle \delta u_x \rangle$
(Fig. \ref{mean_IC_inertia}a) is similar for all particle sets: profiles appear
rather flat in the central region of the channel ($60 < z^+ < 150$) where
$\langle \delta u_x \rangle$ is nearly zero and develop a
peak in the near-wall region where $\langle \delta u_x \rangle$
becomes more and more negative as the wall is approached.
Negative $\langle \delta u_x \rangle$ means over-estimation of the
streamwise drag acting on the particles due to filtering.
Maximum overestimation occurs at $z^+ \simeq 15-20$: the
location of this maximum does not change for different inertia.
Inertia has a strong
effect on the magnitude of the peak, which is maximum for $St=5$
and falls off on either side of this value.
This non-monotonic dependence can be explained considering that inertial
particles are
low-pass filters that respond selectively to
removal of sub-grid flow scales according to
a characteristic frequency proportional to
$1/\tau_p$ \cite{Marchioli2008}.
When the frequency of the
removed scales is equal to the characteristic frequency of the
particles, particle response is maximized.
This is precisely the case of Fig. \ref{mean_IC_inertia}a),
in which filtering with CF=4
are observed to produce larger errors for the $St=5$ particles compared to the other
particle sets.
We remark that inertia also plays a role in transferring filtering
errors on fluid velocity into filtering errors on drag, which
are proportional to $\langle \delta {\bf u} \rangle \cdot St$.

Filtering has no significant effect in mean on the spanwise component
$\langle \delta u_y \rangle$ (Fig. \ref{mean_IC_inertia}b) which is
characterized by very small values \ld ${\cal O}(10^{-3})$ \ld oscillating
around zero along the entire channel height.
In the spanwise direction, however, both $\langle {\bar{u}}_s \rangle$
and $\langle u_s \rangle$ attain negligible values.
More interesting observations can be drawn examining the wall-normal component,
$\langle \delta u_z \rangle$ (Fig. \ref{mean_IC_inertia}c).
First, we notice that $\langle \delta u_z \rangle$ profiles develop a
positive peak in the near-wall region which increases in magnitude
with the Stokes number. Moving away from this region, values become
negative and then relax to zero toward the center of the channel.
This means that drag may be either underestimated or overestimated
depending on the particle-to-wall distance, a feature not trivial
to be reproduced in a model.
The observed behavior of $\langle \delta u_z \rangle$ also shows
that filtering in the homogeneous directions has consequences
on the unfiltered inhomogeneous direction.
The filtering error propagates in the wall-normal
direction according to the relation
$
\bar{u}_z ({\bf x},t)= FT^{-1}
\left[
{\hat G}(\kappa_1) \cdot {\hat G}(\kappa_2) \cdot
\hat{u}_z (\kappa_1,\kappa_2,z,t)
\right]
$,
see Eq.  (\ref{cut-off}).

We remark that all components of $\delta {\bf u}$ would be rigorously equal to zero
if computed at fixed grid points rather than along particle trajectories.
This because filtering does not affect the mean value of the velocity.
In a-priori and a-posteriori tests, it is customary to analyze the effects
of filtering
on the LES fluid velocity fields at fixed points (see e.g.
\cite{kv05,Marchioli2008}).
When dealing with LPT, however, a more
natural approach is to evaluate filtering effects
on the fluid velocity seen by the particles: in this way,
both filtering effects on turbulent structures
and preferential concentration effects are
reproduced in the behavior
of $\mathbf{\delta u}$.
From a physical viewpoint, this establishes a direct link
between $\mathbf{\delta u}$ and the phenomenology of particle
dispersion in turbulent boundary layer.
To elaborate, it is well known that near-wall turbulence is characterized
by the presence of low-speed and high-speed streaks and
that inertial particles preferentially sample low-speed streaks
\cite{Picciotto2005,Picciotto2005b,Soldati2009}.
Visual evidence of this tendency is provided in Fig. \ref{fig:streaks}a, in
which the instantaneous position of the $St=25$ particles comprised
within a near-wall region $10$ wall units thick is superposed to
fluid streaks on a ($x; y$)-plane close to the wall ($z^+ = 10$).
Fluid streaks are rendered using colored contours of the streamwise
fluid velocity fluctuation $u_x'$ in a DNS flow field.
The effect of filtering becomes
clear upon comparison of Fig. \ref{fig:streaks}a with
Figs. \ref{fig:streaks}b-\ref{fig:streaks}c, which demonstrate how rendering of the streaky
structures deteriorates when using a coarse-grained flow field (with
coarsening factors 4 and 8, in this specific case).
Filtering attenuates the velocity fluctuations and smooths the streaks
thus increasing the fluid velocity
seen by the particles: this is precisely what causes
the negative peaks of $\langle \delta u_x \rangle$ in Fig. \ref{mean_IC_inertia}a.
%
%
The behavior of $\langle \delta u_z \rangle$ can also be explained in terms
of filtering effects on the wall-normal fluid velocity seen by the particles,
shown in Fig. \ref{fluidvel_part} for all particle sets. Lines refer to
DNS profiles ($\langle u_{s,z} \rangle$), symbols to a-priori LES profiles
($\langle \bar{u}_{s,z} \rangle$).
Note that
$\langle \delta u_z \rangle = \langle u_{s,z} - \bar{u}_{s,z} \rangle =
\langle u_{s,z} \rangle - \langle \bar{u}_{s,z} \rangle$.
Even though particles exhibit a mean wallward drift, both $\langle u_{s,z} \rangle$
and $\langle \bar{u}_{s,z} \rangle$ are positive (namely directed away from the wall)
in the near-wall region due to particle preferential concentration
in ejection-like environments, where low-momentum fluid is entrained
off the wall \cite{Picciotto2005}.
Filtering reduces the wall-normal fluid velocity seen by the particles very close
to the wall while increasing it towards the center of the channel. This is due to
smoothing of the near-wall structures and attenuation of the sweep
events, which are characterized by coherent motions of high-momentum
fluid to the wall
(see e.g. Fig. 6 in Marchioli et al.\cite{Marchioli2011}).
 
Examination of $\langle \delta {\bf u} \rangle$ confirms that LES is inaccurate in capturing
the near-wall turbulent structures and in reproducing quantitatively their interaction
with particles. Errors introduced by filtering lead to underestimation of preferential
concentration and wall accumulation, as demonstrated by previous studies
\cite{kv05,k06,Marchioli2008}.
To provide a thorough characterization of these errors, in Figs. \ref{rms_IC_inertia}
and \ref{skew_IC_inertia} we examine the higher order moments.
Analysis of the root mean square (rms) of the filtering error,
shown in Fig. \ref{rms_IC_inertia}, highlights
the reduction of the fluid velocity fluctuations in LES
already found in error analyses carried out at fixed
(Eulerian) points \cite{kv05}. SGS models previously used for LPT in LES fields,
like approximate deconvolution or filtering inversion, are mainly aimed
at counteracting this effect by recovering the correct amount of
fluid and particle velocity fluctuations in the particle equations \cite{k06,Marchioli2008}.
Note that, except in the very-near wall region, rms values of the error are in general
much higher than the mean values along the same flow direction: only in the
streamwise direction and within the buffer layer, $rms(u_x)$
and $\langle  \delta u_x \rangle$ have the same order of magnitude.
No significant effects due to particle inertia are observed.

The skewness and flatness factors of the filtering error in each flow direction
are shown in Fig. \ref{skew_IC_inertia}, indicated as $S(\mathbf{\delta u})$
and $F(\mathbf{\delta u})$ respectively.
As for the rms, the most interesting results are obtained for the skewness
in the streamwise and wall-normal directions ($S(\delta u_x)$
and $S(\delta u_z)$ in Figs. \ref{skew_IC_inertia}a and
\ref{skew_IC_inertia}e, respectively): values of the skewness in the spanwise direction
($S(\delta u_y)$ in Figs. \ref{skew_IC_inertia}c remain very small
along the entire channel height, roughly oscillating around zero.
Qualitatively, the skewness profiles in the streamwise and wall-normal directions
are similar: moving from the wall toward the center of the channel,
both $S(\delta u_x)$ and $S(\delta u_z)$ exhibit
high positive values in the viscous sublayer followed by low negative values
in the buffer region which then relax toward zero at the channel centerline.
This behavior indicates that significant asymmetric deviations from Gaussianity
are expected in the probability distribution functions of the streamwise
and wall-normal components of the filtering error due to significant contributions
given by strong isolated events.
This finding is consistent with previous
analyses \cite{kv05,Marchioli2008} which showed that filtering impacts particle dynamics
mainly through attenuation of low-speed streaks (negative velocity fluctuations)
and ejection events.
Regarding inertial effects, changes in the Stokes number modify the
$S(\delta u_x)$ profiles
particularly in the central portion of the channel,
while larger deviations among the $S(\delta u_z)$
profiles are observed in the near-wall region.
Inertia has almost no effect on the flatness factors, shown in
Figs. \ref{skew_IC_inertia}b,  \ref{skew_IC_inertia}d), and  \ref{skew_IC_inertia}f:
all profiles are characterized by high values at the wall which gradually
decrease to centerline values of about 5 for $F(\delta u_x)$ and
$F(\delta u_z)$ and of about 4 for $F(\delta u_y)$.
Note that these values
are higher than 3, which would be obtained if the error distribution
was Gaussian.

\subsection{Influence of filter type and width on the filtering error}
 \label{statistics_filter}
 
To investigate the effect of filter type and width on the statistical
moments of $\delta {\bf u}$, we focus on the $St=5$ particles.
Filter effects on the mean value of the filtering error
$\langle \delta {\bf u} \rangle$ are shown in Fig. \ref{mean_IC_filter}.
The streamwise component ($\langle \delta u_x \rangle$, Fig. \ref{mean_IC_filter}a),
exhibits a near-wall negative peak whose absolute value increases with the filter width,
namely with CF. This behavior is common to all particle sets investigated in
this study.
Regarding the filter type, it is observed that the top-hat filter introduces
a slightly stronger smoothing of the fluid velocity
fluctuations compared to the cut-off filter, especially for CF=2 and CF=4,
The mean spanwise component of $\langle \delta {\bf u} \rangle$ is not
shown since it is characterized by values that are nearly zero throughout
the channel.
The wall-normal component $\langle \delta u_z \rangle$ is shown in
Figure \ref{mean_IC_filter}b. It is evident that the behavior of
$\langle \delta u_z \rangle$ is influenced significantly
by the filter width rather than by the filter type. As CF increases,
the profiles develop positive and negative peaks of different intensity;
also the distance from the wall at which these peaks occur changes considerably:
all peak locations shift toward the center of the channel, indicating
that filtering errors propagate their effect over a larger portion
of the computational domain.

Analysis of the higher-order moments also shows that effects due to the filter width
are much stronger than those associated with the filter type.
No effect of the filter type is observed on the rms of the filtering error,
reported in Fig. \ref{rms_IC_filter}. Profiles change only with the filter
width, reaching higher values as CF increases because of the larger amount
of fluid velocity fluctuations damped in more coarse-grained fields.
Filter type effects are minor also on the skewness and flatness factors,
shown in Fig. \ref{skw_IC_filter}. When the filter width is increased,
the skewness factor in the streamwise direction (Fig. \ref{skw_IC_filter}a)
increases to high positive values very close to the wall and decreases to
low negative values outside the buffer layer ($z^+ > 30$).
In this outer region of the flow, the skewness in the
wall-normal direction (Fig. \ref{skw_IC_filter}e) increases significantly,
switching from negative to positive values as CF increases.
High positive value of the skewness mean that $\delta {\bf u}$ more
frequently attains large positive rather than negative values,
and of course the reverse for negative skewness.
The largest filter width effects on the flatness factor are observed
in the streamwise direction (Fig. \ref{skw_IC_filter}b)
and produce a decrease of $F(\delta u_x)$
throughout the entire channel: for small and intermediate
widths ($CF=2$ and 4) the flatness factor is everywhere higher
than 3, for large widths (e.g. $CF=8$) values fall below 3
in the region $10 < z^+ < 30$.
We remind that high values of the flatness factor, like those
characterizing the spanwise direction (Fig. \ref{skw_IC_filter}d),
and the wall-normal direction (Fig. \ref{skw_IC_filter}f), indicate
that error fluctuations are often larger than the variance of the
distribution and that error fluctuations have an intermittent
character.
%
%

\section{Probability distribution of the filtering error}
\label{PDF}
In the previous section, we have characterized statistically a lower bound for the error produced by the filtering procedure inherent to the LES approach.
This error is intrinsic of LPT in LES flow fields and its mean and higher
order moments depend both on particle inertia and on filter width.
In this section, we focus on the characterization of the one-point Eulerian
probability density function (PDF) of the filtering error.
This function can be derived from the more general Lagrangian PDF and
is the most practical for direct evaluation of the errors made in the
calculation of the statistical moments\cite{Pope_2,Min_01}.
In the frame of the present
study, the issues of interest about the PDF of the filtering error can be
summarized by the following questions:
\begin{enumerate}
\item Is the filtering error deterministic or stochastic?
\item Does it have a simple distribution, for instance Gaussian stochastic?
\item Is it universal or flow-dependent ?
\end{enumerate}
To answer these questions, we consider the PDFs of the streamwise and
wall-normal component of the lower-bound filtering error $\delta {\bf u}$.
These quantities, indicated hereinafter as $ p(\delta u_x) $ and
$ p(\delta u_z) $ respectively, have been
computed averaging over the same fluid slabs defined in Sec. \ref{statistics}
and over the same time interval given in  Sec. \ref{IC}.
%
%
In addition, PDFs have been normalized and shifted to be centered around zero:  
without shifting the profiles, both $ p(\delta u_x) $ and
$ p(\delta u_z) $ would be centered around different
values of $\delta u_x$ and $\delta u_z$ for the different particle sets,
as a result of the dependence of the filtering error on particle inertia
(see Sec. \ref{statistics_inertia}).
Probability distributions in the spanwise direction are not shown
as they are very similar to those obtained for $ p(\delta u_z) $
and add very little to the discussion.

We examine first the PDF behavior at varying filter widths,
shown in Figs. \ref{pdf_x_co} and \ref{pdf_z_co}.
To this aim, we focus on results obtained for the $St=5$ particles
when the cut-off filter is used (as for the statistical moments
of $\delta {\bf u}$, effects on the PDF behavior due to the filter
type appear negligible and hence will not be discussed).
In these figures
profiles relative to four different distances from the wall are
considered to emphasize flow anisotropy effects: clearly, choosing
a non-homogeneous anisotropic flow as reference configuration 
is of particular importance to devise general conclusions.
For comparison purposes, the normalized Gaussian distributions (lines)
with variance equal to that of the computed PDFs (symbols) at the
same wall distance are also shown.
A first (expected) result is that the PDF variance increases with the
filter width in all flow directions and at all wall-normal locations
examined.
Another important finding is that, in all cases examined, the error is
not deterministic, displays very complex features and covers a wide range of values
(typical of probabilistic process) in all cases.
Specifically, the PDFs are not Gaussian and their behavior varies strongly
with the distance from the wall. This observation is not obvious since
the error behavior might be quite different and even perfectly Gaussian
in spite of the non-homogeneous nature of the
flow.
Figs. \ref{pdf_x_co} and \ref{pdf_z_co} show that both
$ p(\delta u_x) $ and $ p(\delta u_z) $ are
characterized by larger flatness than the corresponding Gaussian distributions
for all filter widths and at almost all wall-normal locations.
Furthermore, both PDFs are skewed near the wall \ld
at $z^+ \simeq 4$, Figs. \ref{pdf_x_co}a and \ref{pdf_z_co}a \ld and
outside the buffer region \ld at $z^+ \simeq 65$, Figs. \ref{pdf_x_co}d and \ref{pdf_z_co}d.
This is again related to poor description of coherent structures, e.g.
low-speed streaks and turbulent vortices.
At intermediate wall-normal locations, PDFs exhibit a nearly Gaussian behavior
only at $z^+ \simeq 17$ for all filter widths but the largest
\ld Figs. \ref{pdf_x_co}b and \ref{pdf_z_co}b;
at $z^+ \simeq 37$,
where Reynolds stresses reach a maximum, the non-Gaussian behavior of the PDF
profiles increases again \ld Figs. \ref{pdf_x_co}c and \ref{pdf_z_co}c.
It can also be observed that
(i) $ p(\delta u_z) $ profiles become
flatter at increasing filter widths (Fig. \ref{pdf_z_co}), and
%
(ii) filter width effects lead to higher skewness and mildly
decreasing flatness of the PDF,
consistently with results of Figs. \ref{skw_IC_filter}a and \ref{skw_IC_filter}b.
For the largest filter width (CF=8), there is a change from positive skewness in
the very near wall region ($z^+ \simeq 4$) to negative skewness away from the wall.

Additional information on $ p(\delta u_x) $ and
$ p(\delta u_z) $ can be obtained examining how these
quantities are modified by particle inertia.
To this aim, we focus on results obtained using a cut-off filter with CF=4,
shown in Figs. \ref{pdf_x_CF4} and \ref{pdf_z_CF4}.
%
A first observation is that the effect of inertia on the shape of the PDFs is
almost negligible, as we have tried to emphasize by centering
all profiles around $\delta u_i=0$, where $i=x,z$.
%
A general feature of the PDFs reported in Figs. \ref{pdf_x_CF4} and \ref{pdf_z_CF4}
is that they exhibit higher tails compared to the corresponding Gaussian distributions,
even near the channel center.
This is in line with the analysis of the flatness profiles carried out in Sec.
\ref{statistics_inertia}.
The flatness is larger for the wall-normal probability distributions and
varies mildly with the distance from the wall, the only exception being
the viscous sublayer (see profiles at $z^+ \simeq 4$).
%
In terms of relative importance of the PDF tails,
$ p(\delta u_x) $ profiles exhibit
asymmetries towards negative values of $\delta u_x$ which become particularly
evident outside the viscous sublayer ($z^+ \simeq 17$, 37 and 65);
whereas $ p(\delta u_z) $ profiles are all symmetric with
just a slight asymmetry observed at $z^+ \simeq 17$.
These findings are in agreement with the skewness profiles shown
in Figure \ref{skew_IC_inertia}.

Some general conclusions can be drawn from collective analysis of Figs. \ref{pdf_x_co}
to \ref{pdf_z_CF4}. First, intrinsic filtering errors of LES in LPT can be
described as a stochastic process almost everywhere non-Gaussian with important
and long tails (namely large flatness) and appreciable skewness. This process
is significantly influenced by flow anisotropy as demonstrated by the dependency
of skewness of $ p(\delta u_x) $ on the distance from the wall.
This result is important from the modeling viewpoint, as it implies that
neither deterministic models nor stochastic homogeneous models have the capability
to correct fully 
intrinsic error made in the LES approach due to filtering.
Second, the model should adapt dynamically over space to account for local
inhomogeneity and anisotropy of the flow and to
reproduce the correct crossing trajectory effect:
this latter effect, in
particular, is due to the existence of a mean relative
particle-to-fluid velocity (rather than an instantaneous one)
associated with the lack of correlation
between the fluid velocity
seen by the inertial particles and the velocity of fluid particles
\cite{Min_04}.
Finally, results demonstrate that a mandatory requirement to scrutinize model
performance is the capability to reproduce
correctly all spatial turbulent structures (or at least their statistical effect).

\section{Concluding Remarks}
\label{conclusions}

In this work, we proposed a simple procedure to characterize the filtering error
intrinsic of Lagrangian particle tracking in LES flow fields. Specifically, we
focused on the error purely due to filtering of the fluid velocity fields seen
by the particles, for which we quantified a lower bound. This ``minimum'' error
can not be avoided
but potentially can be corrected.
To this aim, we considered an ideal situation in which the exact dynamics of
the resolved velocity field is available and time accumulation of the pure
filtering error on particle trajectories is eliminated a-priori.
%
Evaluation of the statistical properties of the filtering error
has been performed applying filters
of different type and different widths, corresponding roughly
to varying amounts of resolved flow energy, and considering
heavy particles with different inertia.
In particular, mean and higher-order moments as well as one-point
probability distribution functions have been analyzed to provide
information about the key features required in SGS
models for Lagrangian particle tracking to compensate for such error.

The first interesting result is that the mean value of the filtering error
components along particle trajectories is not zero, as it would be obtained
performing an Eulerian analysis of filtering effects at fixed points.
Non-zero mean values result from filtering effects on the coherent turbulent
structures which characterize the near-wall region and are preferentially
sampled by the inertial particles.
Smoothing of the low-speed streaks due to filtering
is associated with a near-wall negative peak of the mean filtering error in the
streamwise direction.
Filtering effects on the near-wall vortical structures, and in turn on sweep
and ejection events, lead to even more complicated mean profiles in the wall-normal
direction: in this case, a near-wall positive peak followed by a negative peak farther
from the wall is observed.
In both directions, particle inertia and filter width change significantly the
qualitative behavior of the error.
Mean value analysis confirms that the error introduced by filtering on inertial
particle motion is not just yield by the reduction of fluid velocity
fluctuations, a well-known effect pointed out in Eulerian error analyses in LES.
This supports the conclusion, drawn in our previous works\cite{Marchioli2008,Marchioli2008b},
that SGS model for the particle equations can not ensure accurate prediction of
particle preferential concentration and near-wall accumulation by only
reintroducing the correct amount of fluid velocity fluctuations.
Improved predictions of LES applied to turbulent dispersed flow necessitate
suitable benchmarks against statistical information at the sub-grid level,
and the present study establishes precisely the minimum 
benchmarking requirements.

Joint analysis of the the higher-order statistical moments and of the
probability distribution functions shows that the filtering error distribution
differs significantly from a normalized Gaussian distribution with the
same variance. All error components are characterized by a flatness
generally larger than the Gaussian value of 3. The streamwise and wall-normal
components are also characterized by skewed distributions, not observed
in the spanwise direction: both positive and negative skewness values
can be attained depending on the distance from the wall.
This indicates that error behavior is intermittent and strongly
affected by flow anisotropy.
Changes in the probability distributions are also found at varying filter
width. Besides the expected increase of the variance of the PDF when
the filter become coarser, noticeable effects are observed on the skewness
and flatness of the streamwise PDF and on the skewness of the wall-normal PDF.
No significant effects are observed when either filter type or particle inertia
are changed.

Present results emphasize the stochastic nature of the pure filtering error.
In our opinion, a possible way to correct this error could be represented
by a generalized stochastic Lagrangian model \cite{pope,Min_01}
in which the statistical {\it signature}
of coherent turbulent structures (e.g. low-speed streaks, sweeps and ejections)
on particle dynamics is accounted for explicitly. In this perspective, the filtering
error characterization and quantification provided by the present study represents
a valuable tool to understand how these structures are affected by filtering and
how sub-grid scale fluid velocities impact the force driving the particles.
Models of this type have already been developed in the context of RANS methods
\cite{Min_04}, followed by extensions to heuristic particle deposition
modeling\cite{Chi_08} and by first attempts at explicitly modeling
the statistical effects of structures on particle statistics in near-wall
regions\cite{Gui_08}.
Much work remains to be done. The first necessary step is of course
to extend the formalism
to the LES framework. To this aim, a detailed Lagrangian analysis of the
error should be performed to complement present results and assist in
model development.

\section*{Acknowledgments}
CINECA supercomputing center (Bologna, Italy) is gratefully acknowledged for generous allowance of computer resources. The authors wish to thank Prof. B.J. Geurts for the interesting discussion which stimulated the present study.

\bibliographystyle{plainnat}

\newpage
\begin{center}
\textbf{List of Tables}
\end{center}
\noindent Table \ref{table:part}:
Particle parameters; $\tau_p$ is the particle relaxation time, $d_p$ the particle
diameter, $V_s$ the particle settling velocity and $Re_p$ the particle Reynolds number.
The superscript $+$ denotes the corresponding quantities expressed in wall units.

\clearpage
\newpage

\begin{table}[t]
\begin{minipage}[t]{15.0cm}
\begin{small}
\begin{center}
\begin{tabular}{c c c c c c}
$St$ & $\tau_{p}~(s)$ & $d_{p}^{+}$ & $d_{p}$ (${\mu}m$) & $V_{s}^{+}=g^+\cdot St$  & $Re_{p}^{+}=V_s^+ \cdot d_{p}^{+} / \nu^+$\\
\hline
1   & $1.133 \cdot 10^{-3}$ & $0.153$ & $~20.4$ & $0.0943$ & $0.01443$\\
5   & $5.660 \cdot 10^{-3}$ & $0.342$ & $~45.6$ & $0.4717$ & $0.16132$\\
25  & $28.32 \cdot 10^{-3}$ & $0.765$ & $102.0$ & $2.3584$ & $1.80418$\\
\end{tabular}
\vspace{0.3cm}
\caption{Particle parameters; $\tau_p$ is the particle relaxation time, $d_p$ the particle diameter, $V_s$ the particle settling velocity and $Re_p$ the particle Reynolds number.
The superscript $+$ denotes the corresponding quantities expressed in wall units.}
\label{table:part}
\end{center}
\end{small}
\end{minipage}
\end{table}

\begin{figure}
\begin{tabular}{c}
\centerline{\includegraphics[height=6.cm,width=8.4cm,angle=0.]{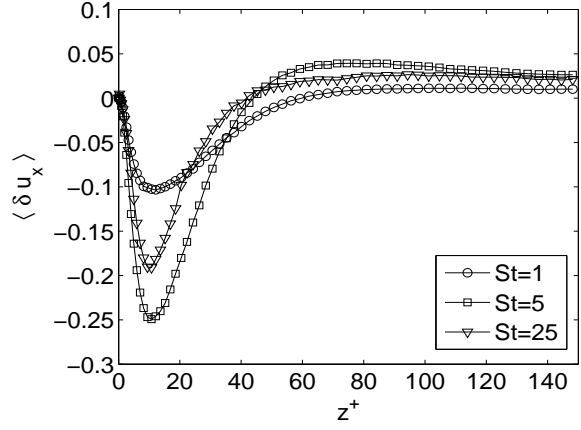}}\\ 
(a)\\
\centerline{\hspace*{0.3cm} \includegraphics[height=6.cm,width=8.cm,angle=0.]{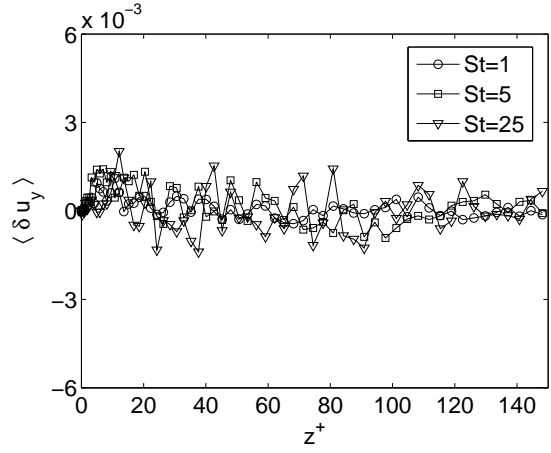}}\\
(b)\\
\centerline{\includegraphics[height=6.cm,width=8.6cm,angle=0.]{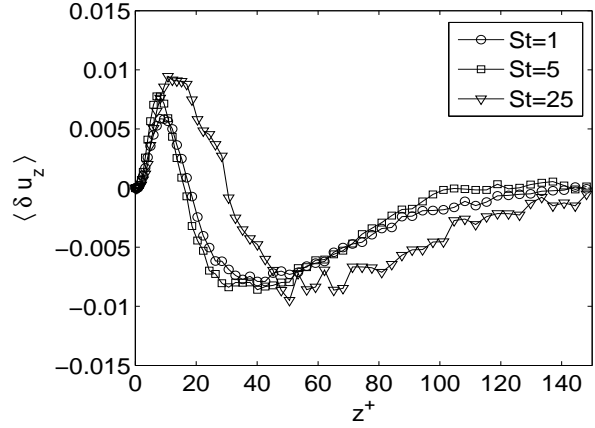}} \\
(c)\\
\end{tabular}
\vspace{0.2cm}
\caption{Mean values of the filtering error $\delta {\bf u}$
in the streamwise (a), spanwise (b) and wall-normal (c) directions
as a function of $z^+$ at varying particle inertia.
Profiles refer to the cut-off filter with CF=4.}
\label{mean_IC_inertia}
\end{figure}

\begin{figure}
\centerline{\includegraphics[height=12.cm,angle=270.]{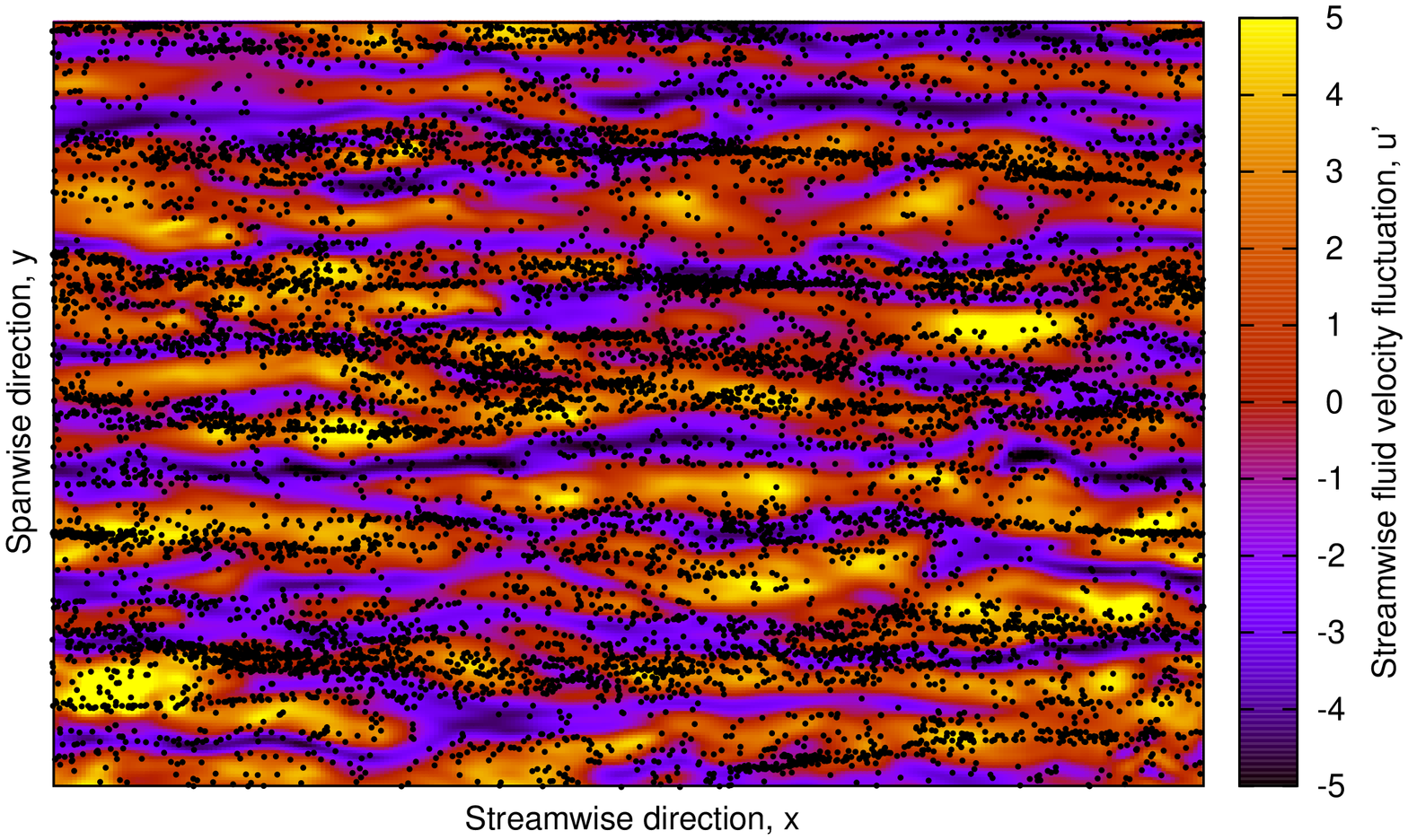}}
\vspace{-6.7cm}
\hspace{-9.5cm}
(a)
\vspace{3.2cm}

\centerline{\includegraphics[height=12.cm,angle=270.]{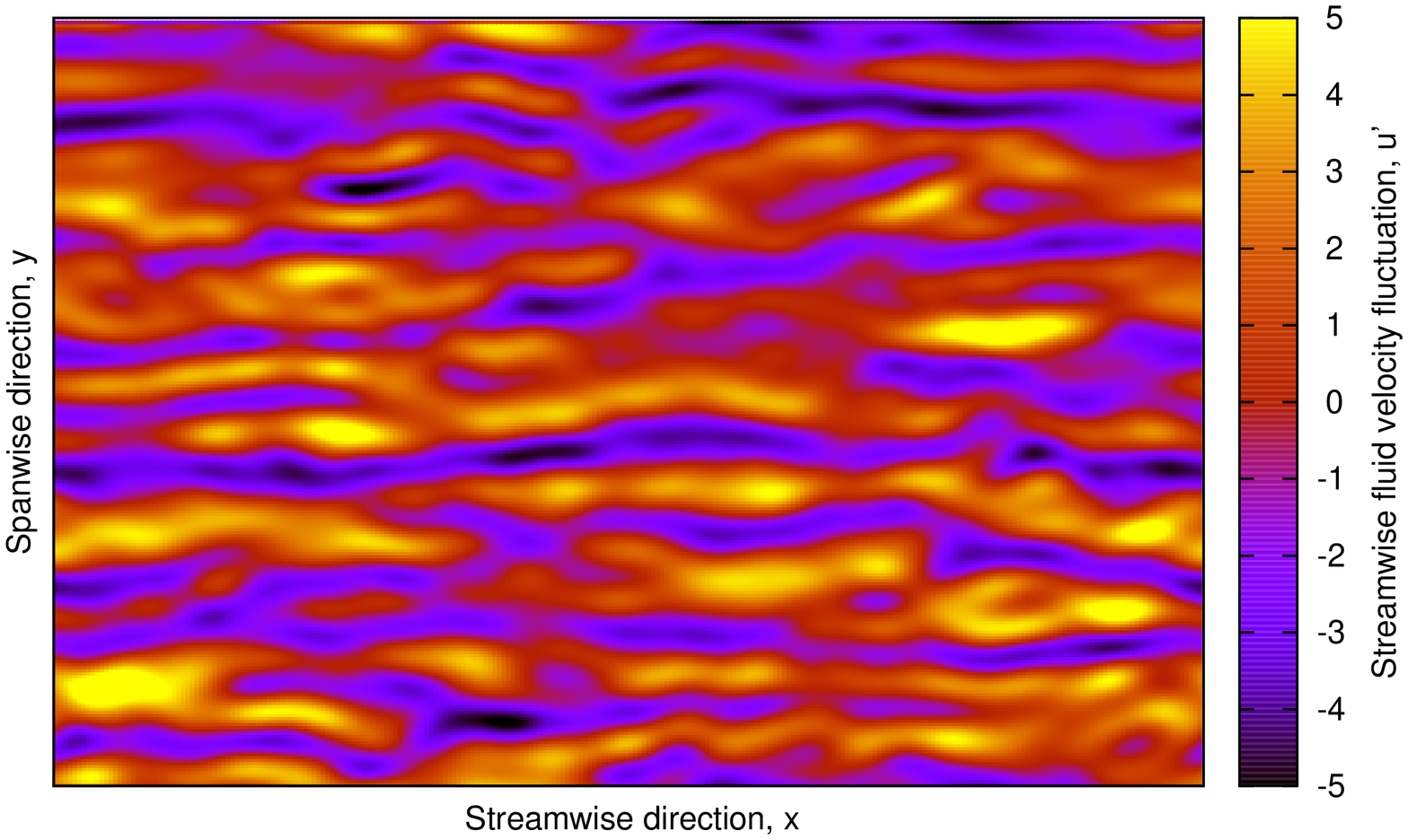}}
\vspace{-6.7cm}
\hspace{-9.5cm}
(b)
\vspace{3.2cm}

\centerline{\includegraphics[height=12.cm,angle=270.]{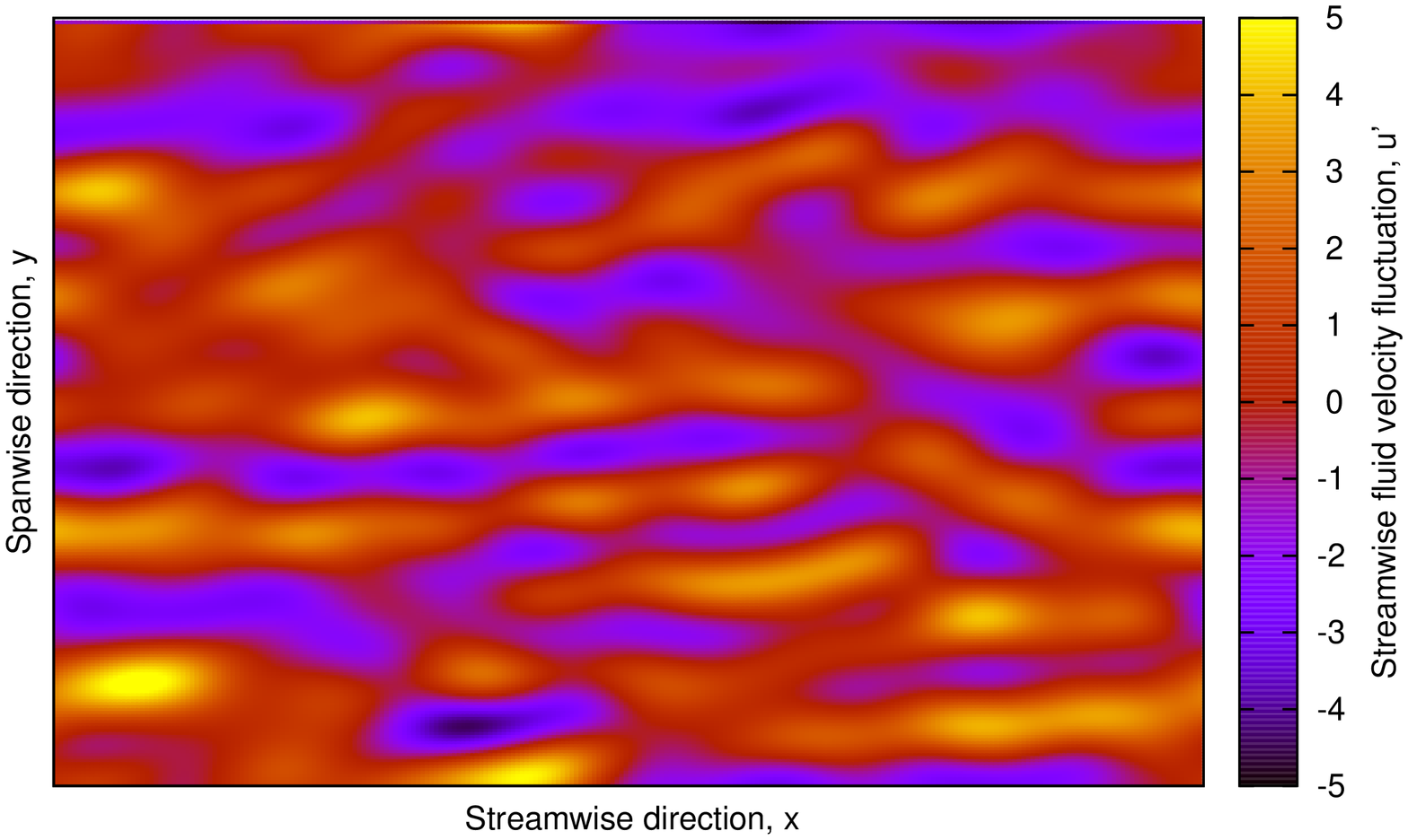}}
\vspace{-6.7cm}
\hspace{-9.5cm}
(c)
\vspace{4.7cm}
\caption{Fluid streaks in the near-wall region. Low-speed (resp. high-speed) streaks
are rendered using colored contours of negative (resp. positive) streamwise fluctuating
velocity $u_x'$ on a horizontal plane at $z^+ = 10$ from the wall.
Panel (a) shows the streaky structures obtained from DNS; the
instantaneous distribution of the $St=25$ particles comprised
between $z^+ = 10$ and the wall is also shown to highlight
preferential distribution into the low-speed streaks.
Panels (b) and (c) show the streaky structures obtained from a-priori
DNS with coarsening-factor 4 and 8, respectively.
}
\label{fig:streaks}
\end{figure}

\begin{figure}
\centerline{\includegraphics[height=6.cm,width=8.5cm,angle=0.]{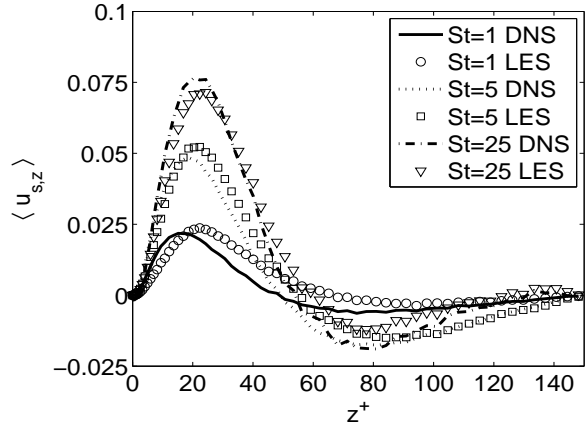}}
\caption{Effects of filtering on the mean wall-normal fluid velocity seen by the
particles at varying particle inertia. Profiles refer to the cut-off filter with CF=4.}
\label{fluidvel_part}
\end{figure}

\begin{figure}
\begin{tabular}{c}
\centerline{\includegraphics[height=6.cm,width=8.5cm,angle=0.]{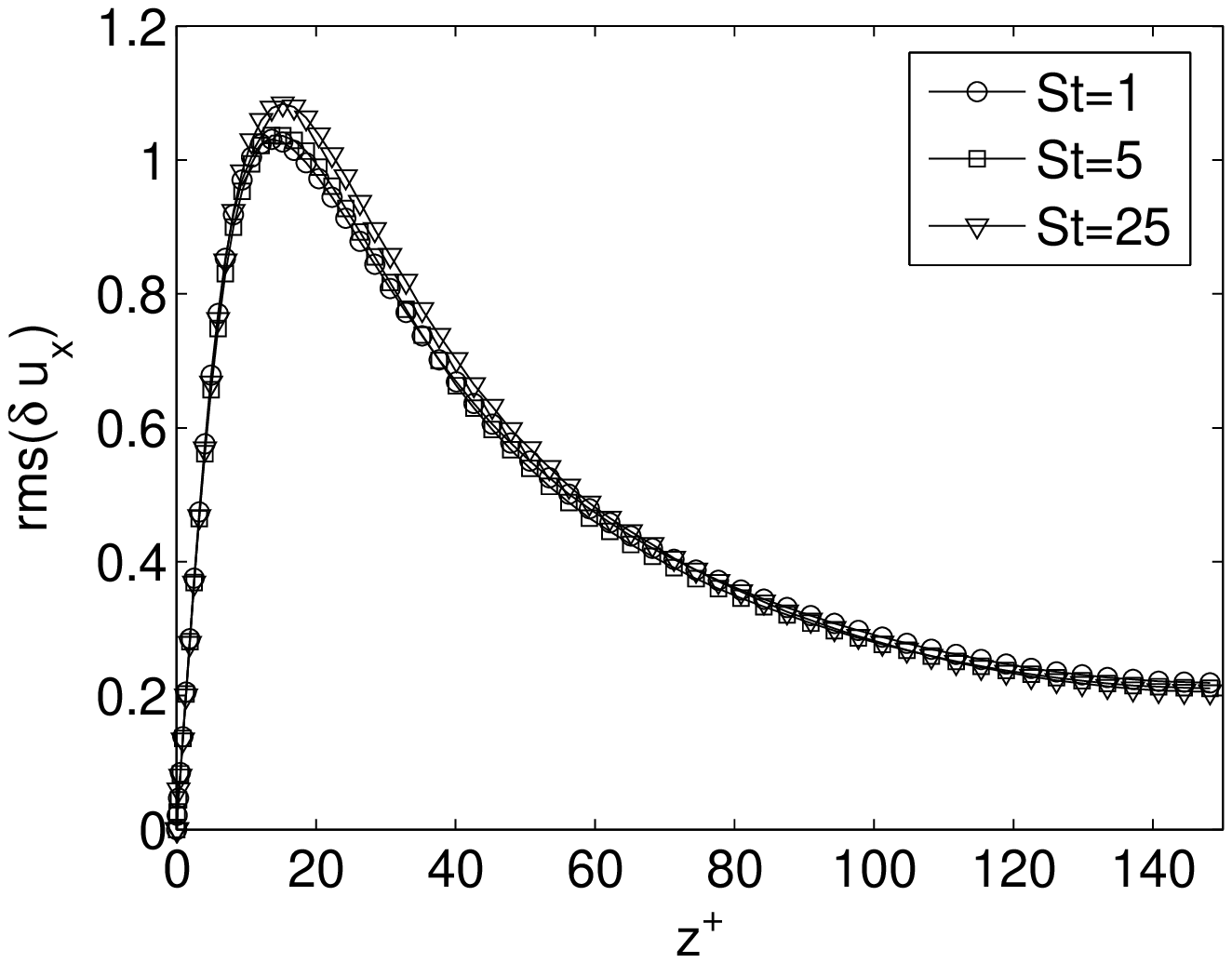}}\\
(a)\\
\centerline{\includegraphics[height=6.cm,width=8.5cm,angle=0.]{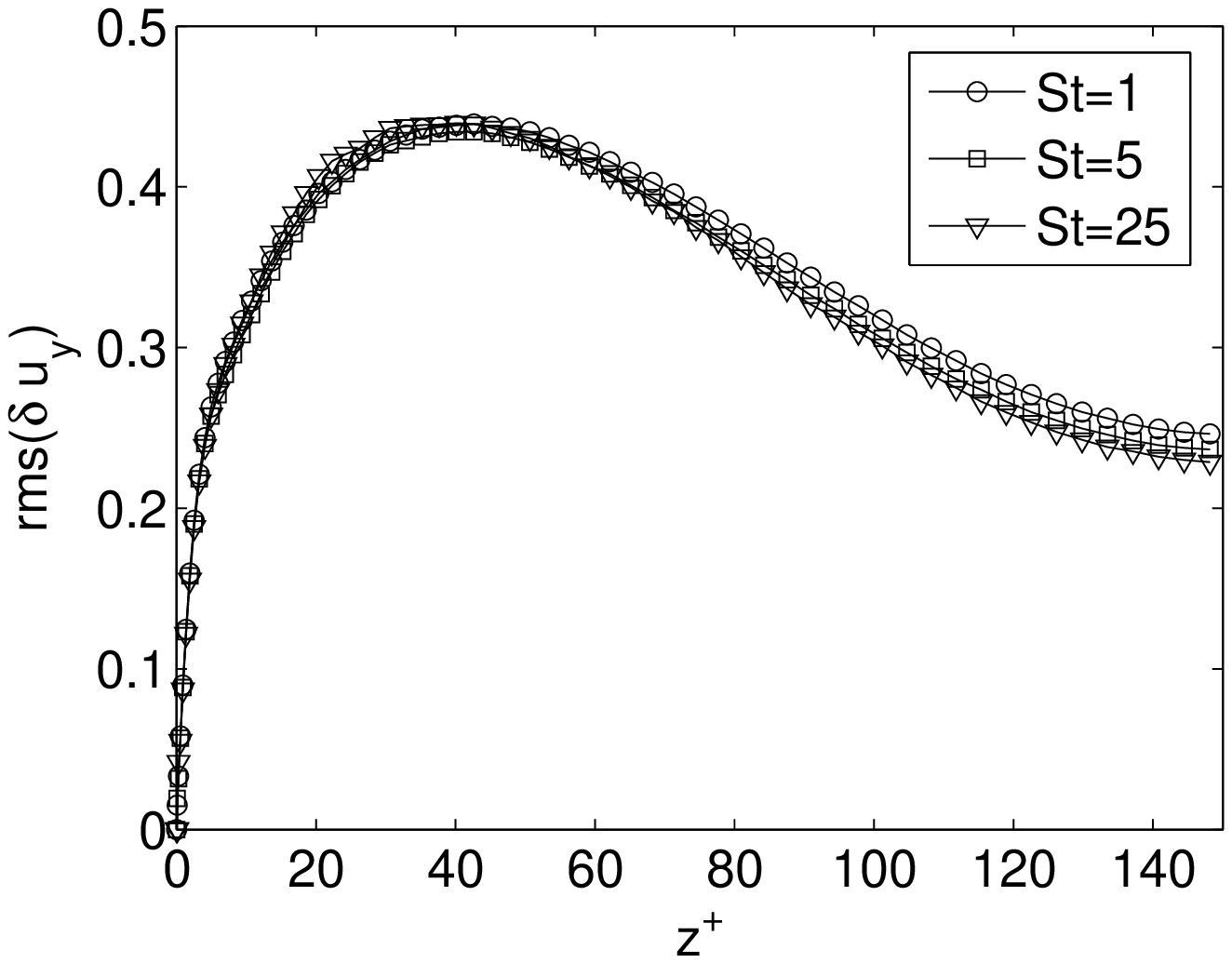}}\\
(b)\\
\centerline{\includegraphics[height=6.cm,width=8.5cm,angle=0.]{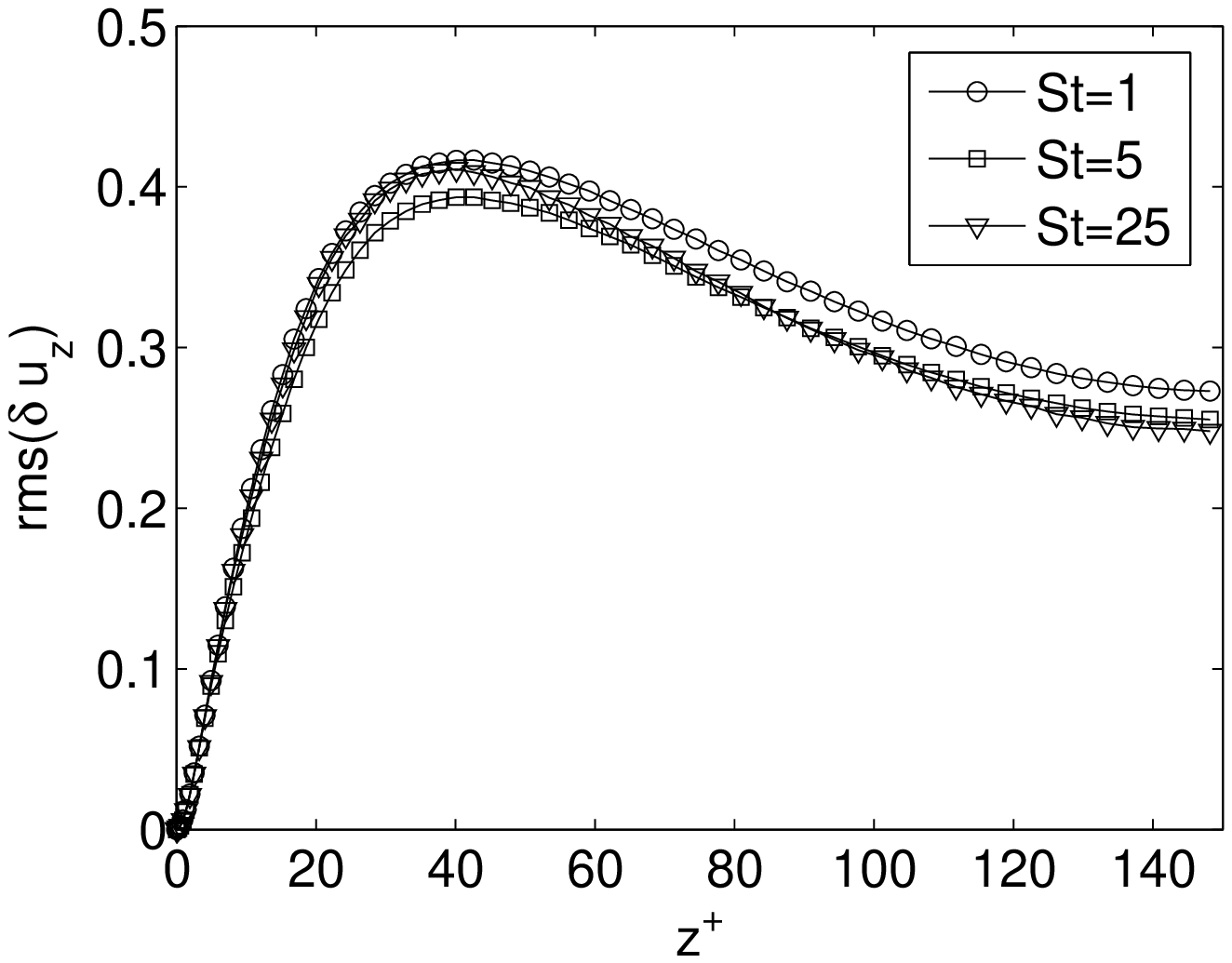}}\\
(c)
\end{tabular}
\vspace{0.2cm}
\caption{Rms of the filtering error $\delta {\bf u}$ in the streamwise (a),
spanwise (b) and wall-normal (c) directions as a function of $z^+$ at
varying particle inertia. Profiles
refer to the cut-off filter with CF=4.}
\label{rms_IC_inertia}
\end{figure}

\begin{figure}
\begin{tabular}{c}
\centerline{\includegraphics[height=6.cm,width=8.cm,angle=0.]{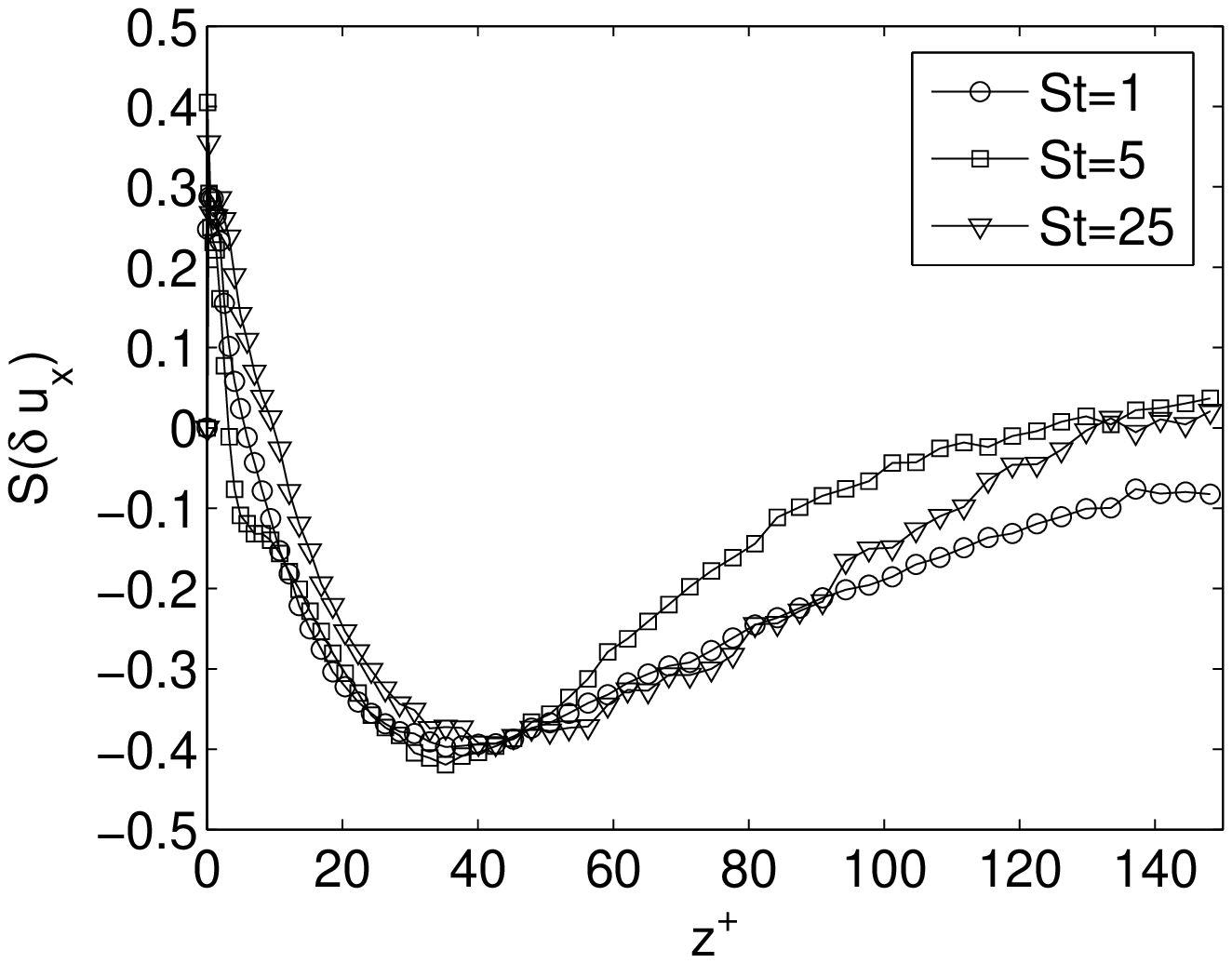}
\includegraphics[height=6.cm,width=8.cm,angle=0.]{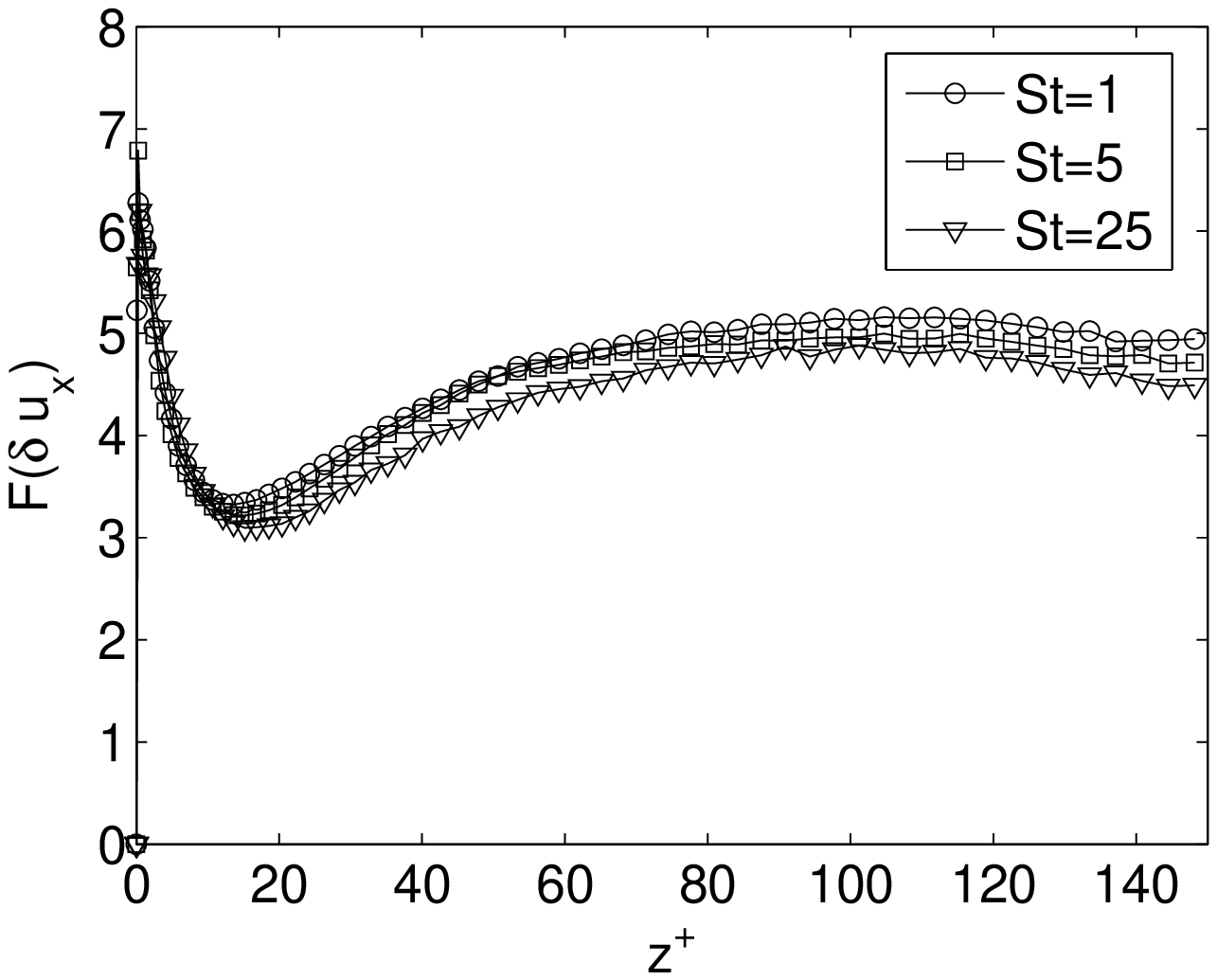}}\\
(a) \hspace{8.0cm} (b)\\
\centerline{\includegraphics[height=6.cm,width=8.cm,angle=0.]{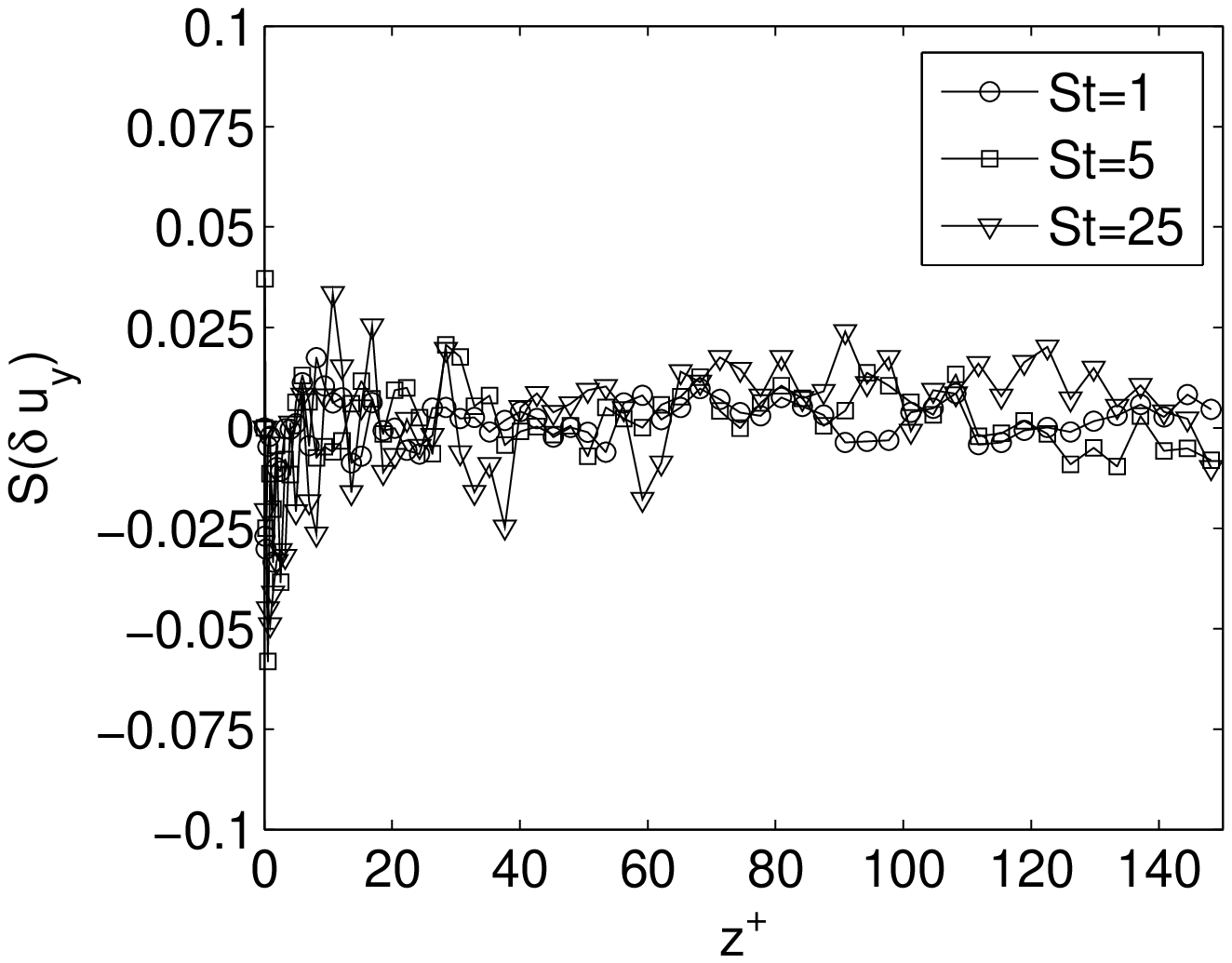}
\includegraphics[height=6.cm,width=8.cm,angle=0.]{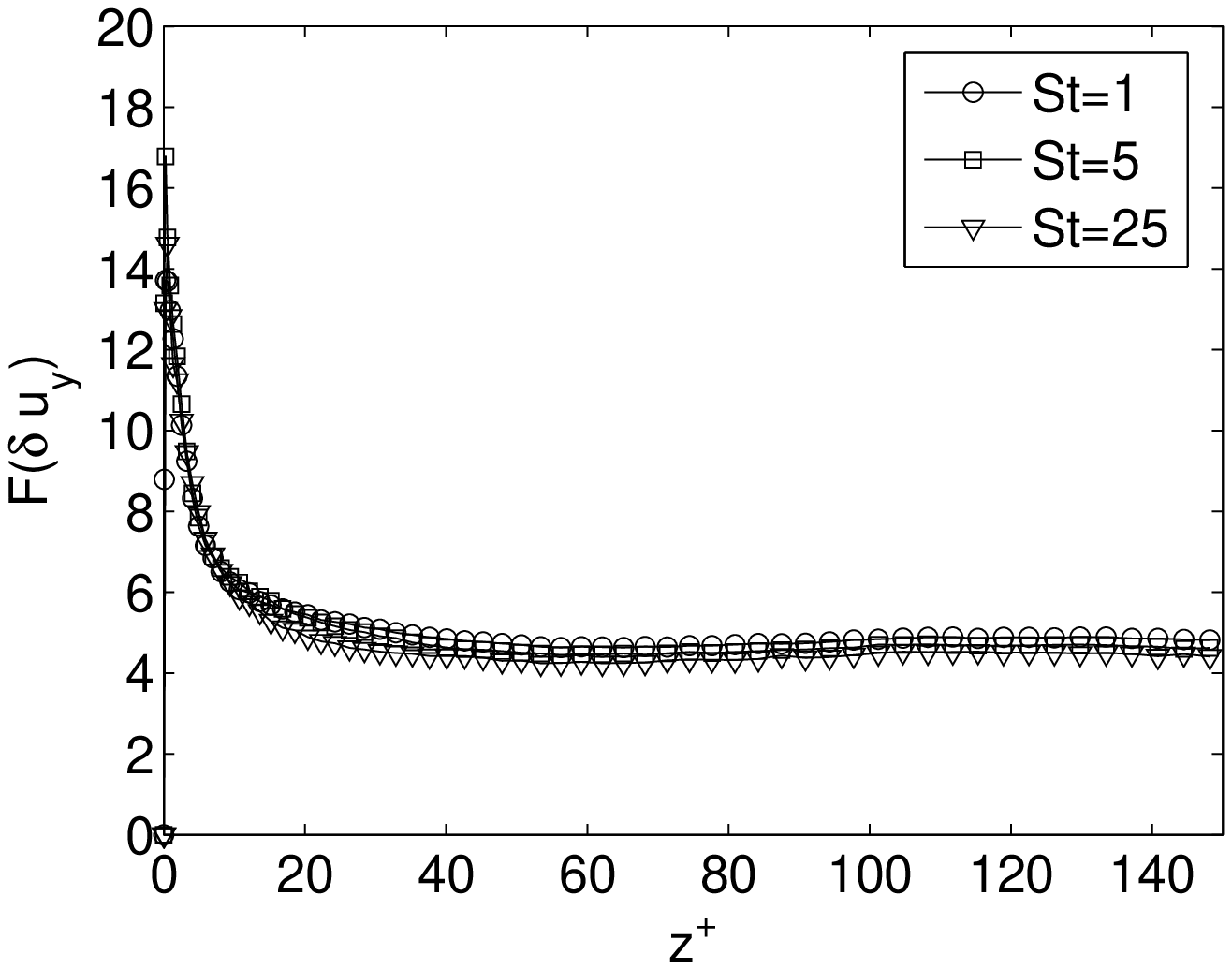}}\\
(c) \hspace{8.0cm} (d)\\
\centerline{\includegraphics[height=6.cm,width=8.cm,angle=0.]{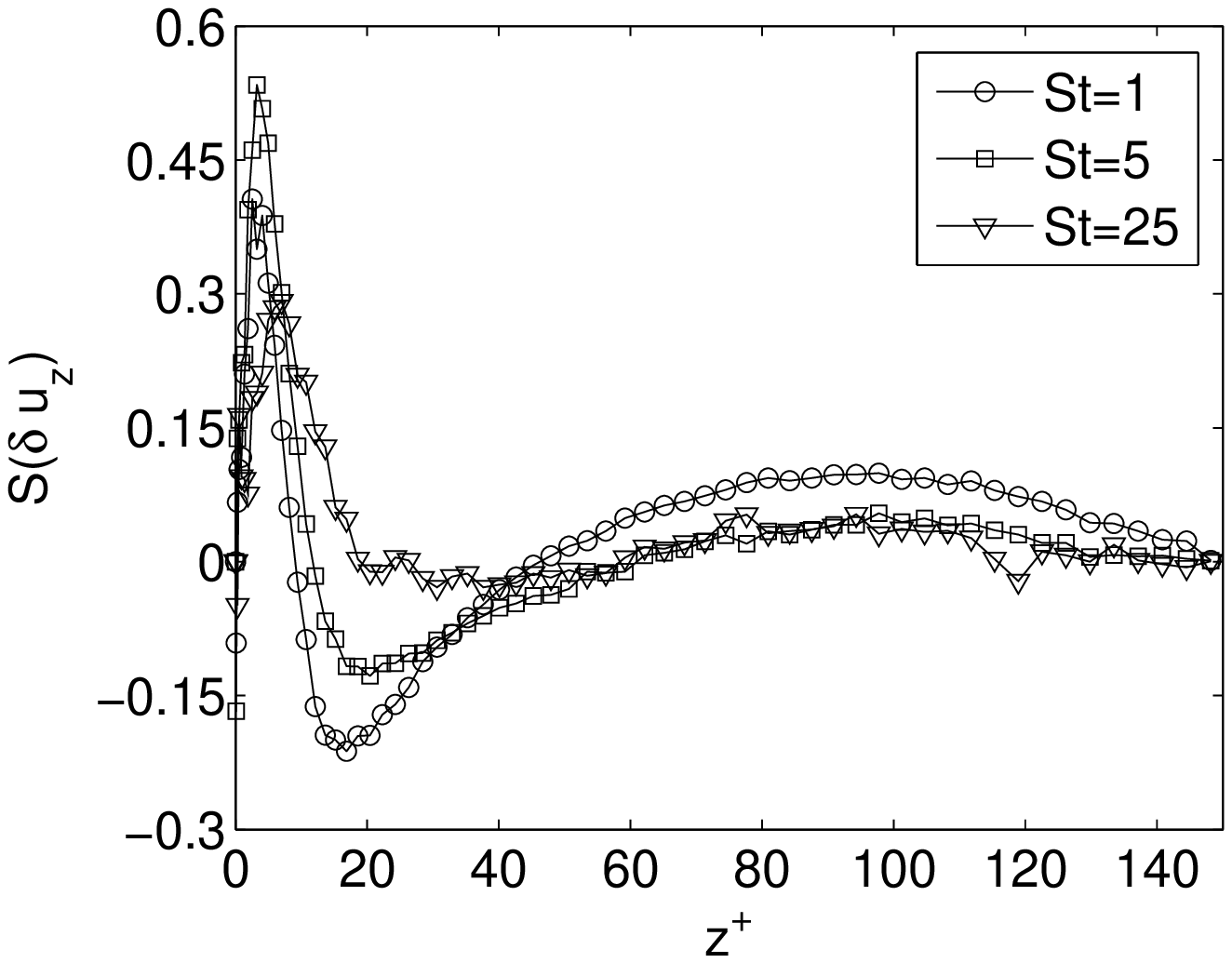}
\includegraphics[height=6.cm,width=8.cm,angle=0.]{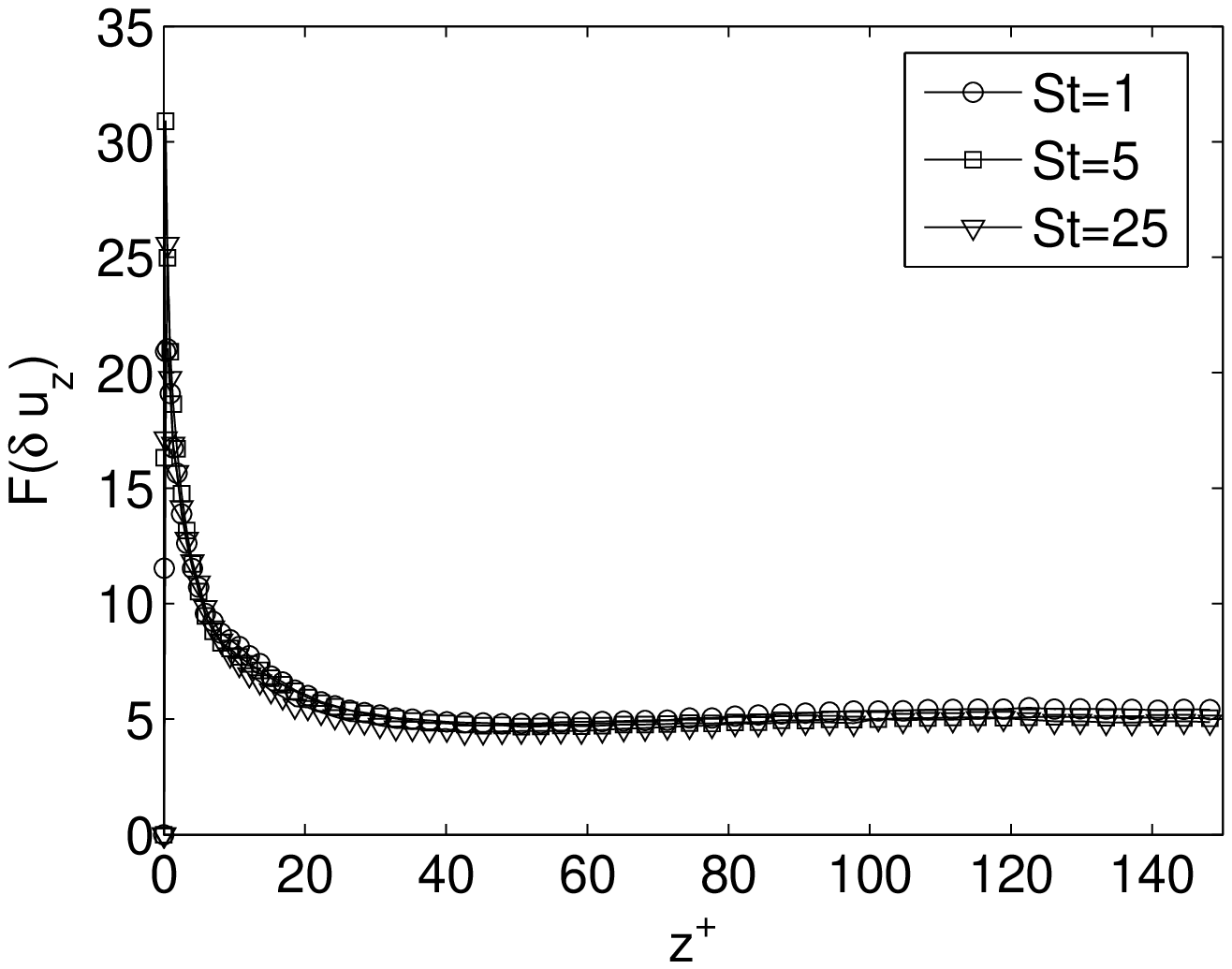}}\\
(e) \hspace{8.0cm} (f)
\end{tabular}
\vspace{0.2cm}
\caption{Skewness and flatness of the filtering error $\delta {\bf u}$
in the streamwise (a-b), spanwise (c-d) and wall-normal (e-f) directions
as a function of $z^+$ at
varying particle inertia. Profiles refer to the cut-off filter with CF=4.}
\label{skew_IC_inertia}
\end{figure}

\begin{figure}
\begin{tabular}{cc}
\includegraphics[width=7.cm,angle=0.]{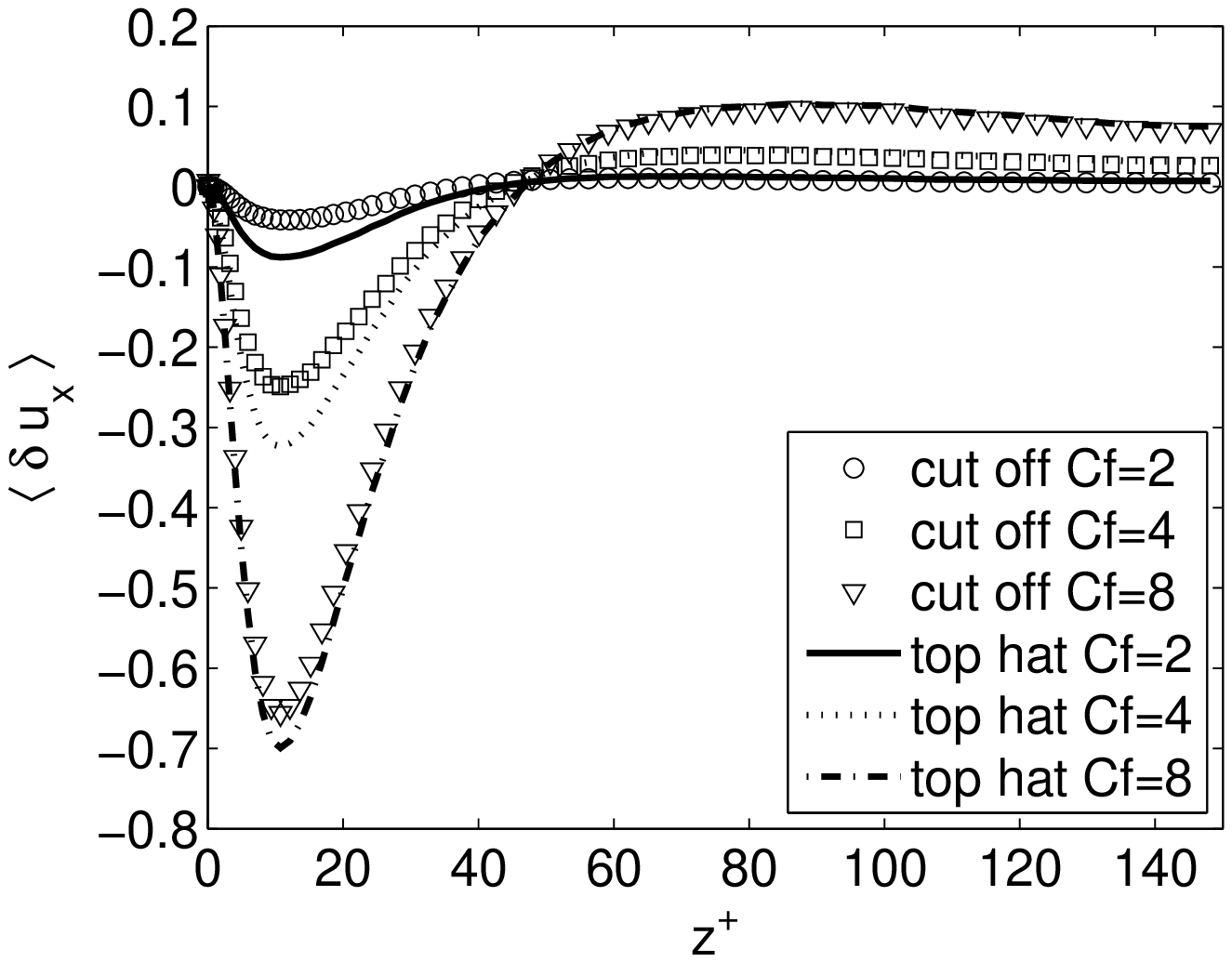} & \includegraphics[width=7.cm,angle=0.]{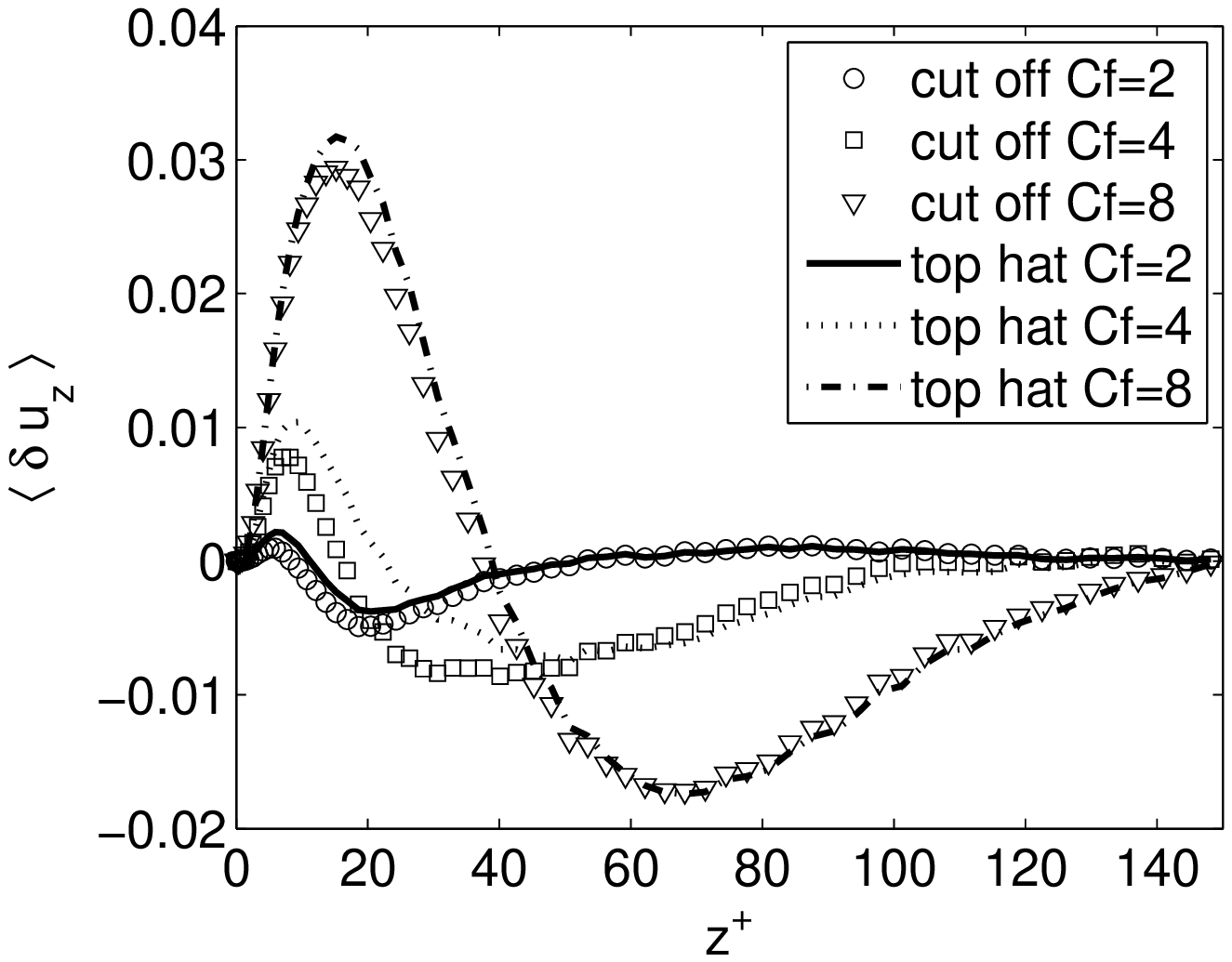}\\
(a) & (b)
\end{tabular}
\vspace{0.2cm}
\caption{Mean values of the filtering error $\delta {\bf u}$
in the streamwise (a), and wall-normal (b) directions
as a function of $z^+$ at varying filter type and width.
Profiles refer to the $St=5$ particles.}
\label{mean_IC_filter}
\end{figure}

\begin{figure}
\begin{tabular}{c}
\centerline{\includegraphics[height=6.cm,width=8.5cm,angle=0.]{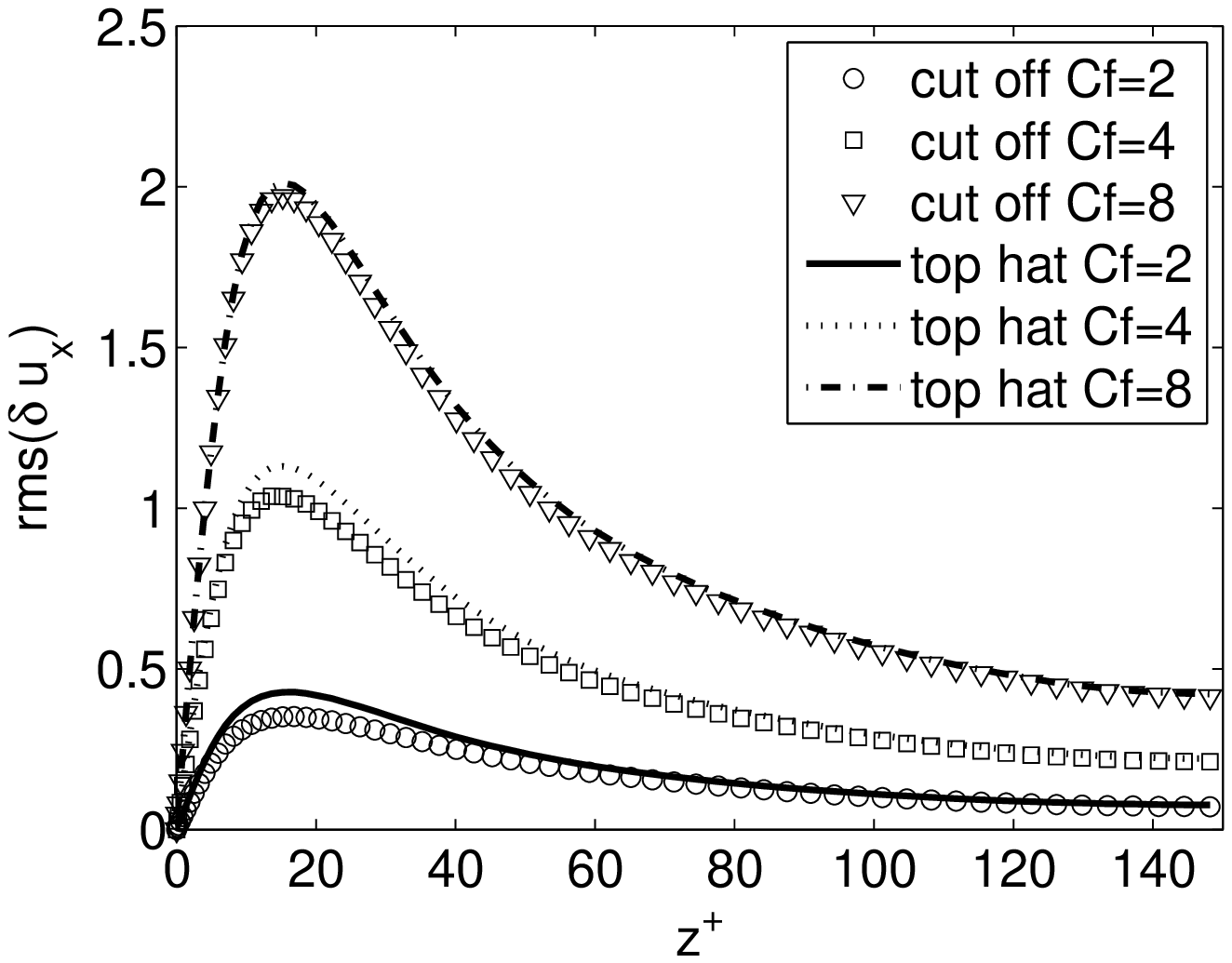}}\\
(a)\\
\centerline{\includegraphics[height=6.cm,width=8.5cm,angle=0.]{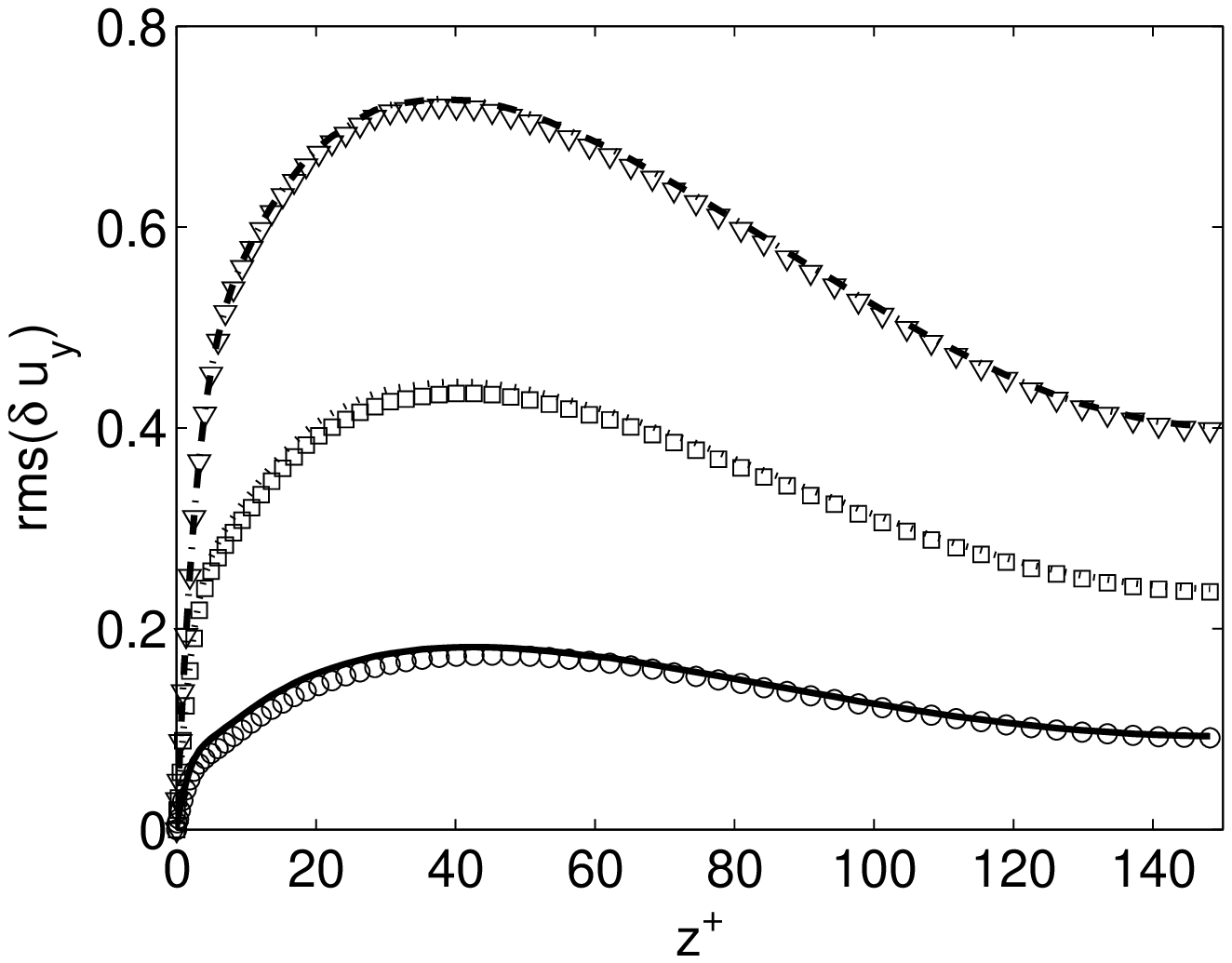}}\\
(b)\\
\centerline{\includegraphics[height=6.cm,width=8.5cm,angle=0.]{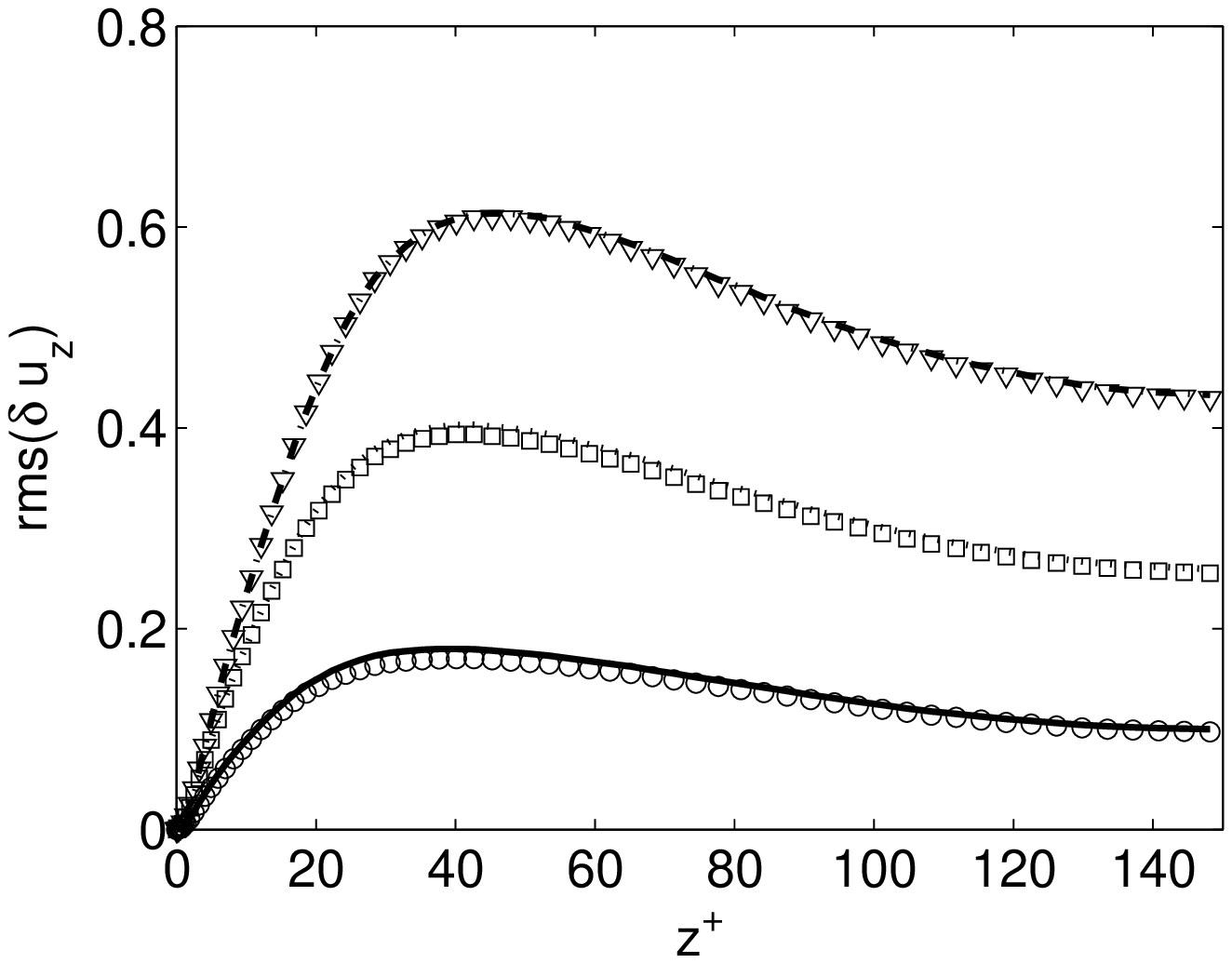}}\\
(c)
\end{tabular}
\vspace{0.2cm}
\caption{Rms of the filtering error $\delta {\bf u}$
in the streamwise (a), spanwise (b) and normal (c) directions
as a function of $z^+$ at varying filter type and width.
Profiles refer to the $St=5$ particles.}
\label{rms_IC_filter}
\end{figure}

\begin{figure}
\begin{tabular}{c}
\centerline{\includegraphics[height=6.cm,width=8.cm,angle=0.]{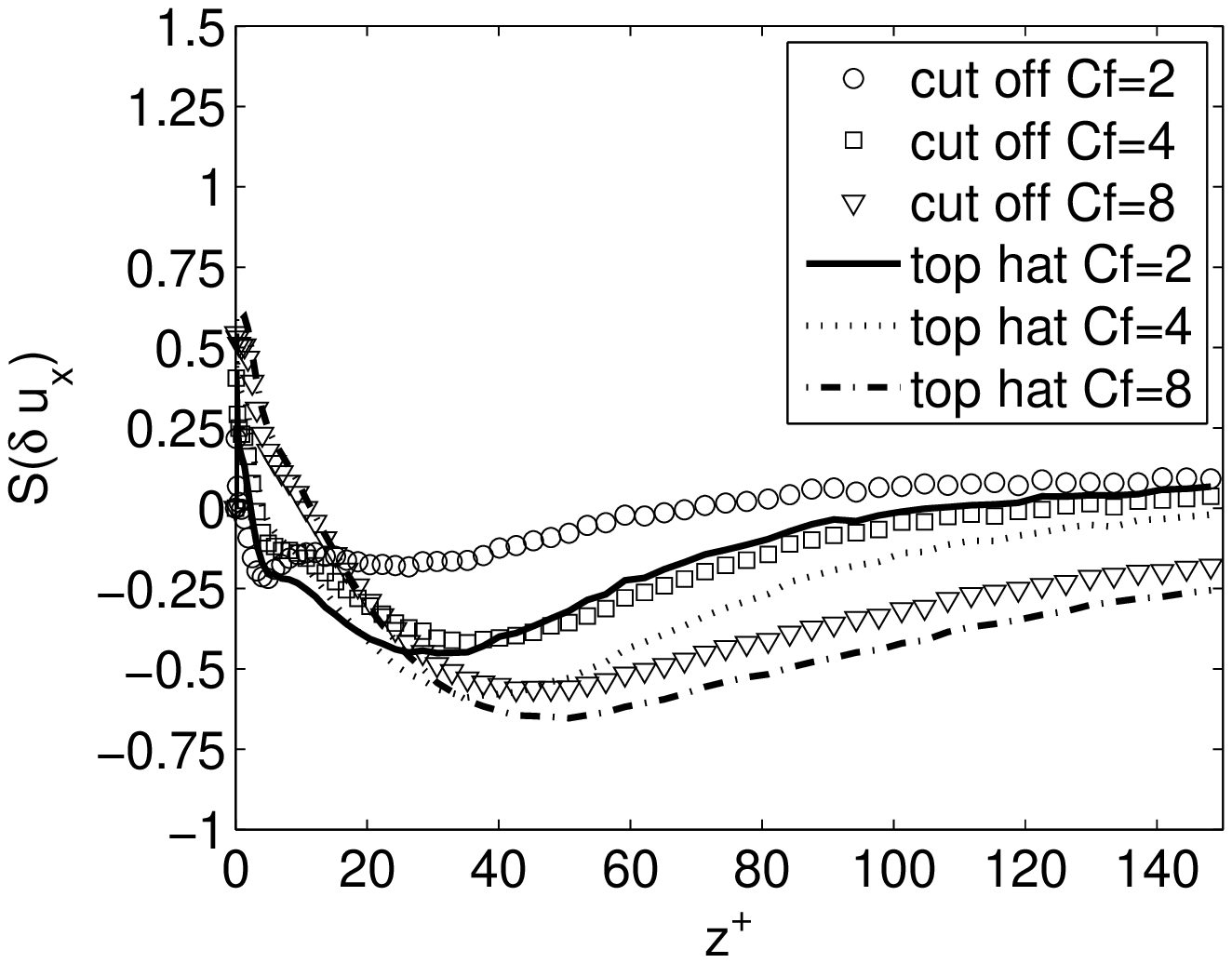}
\includegraphics[height=6.cm,width=8.cm,angle=0.]{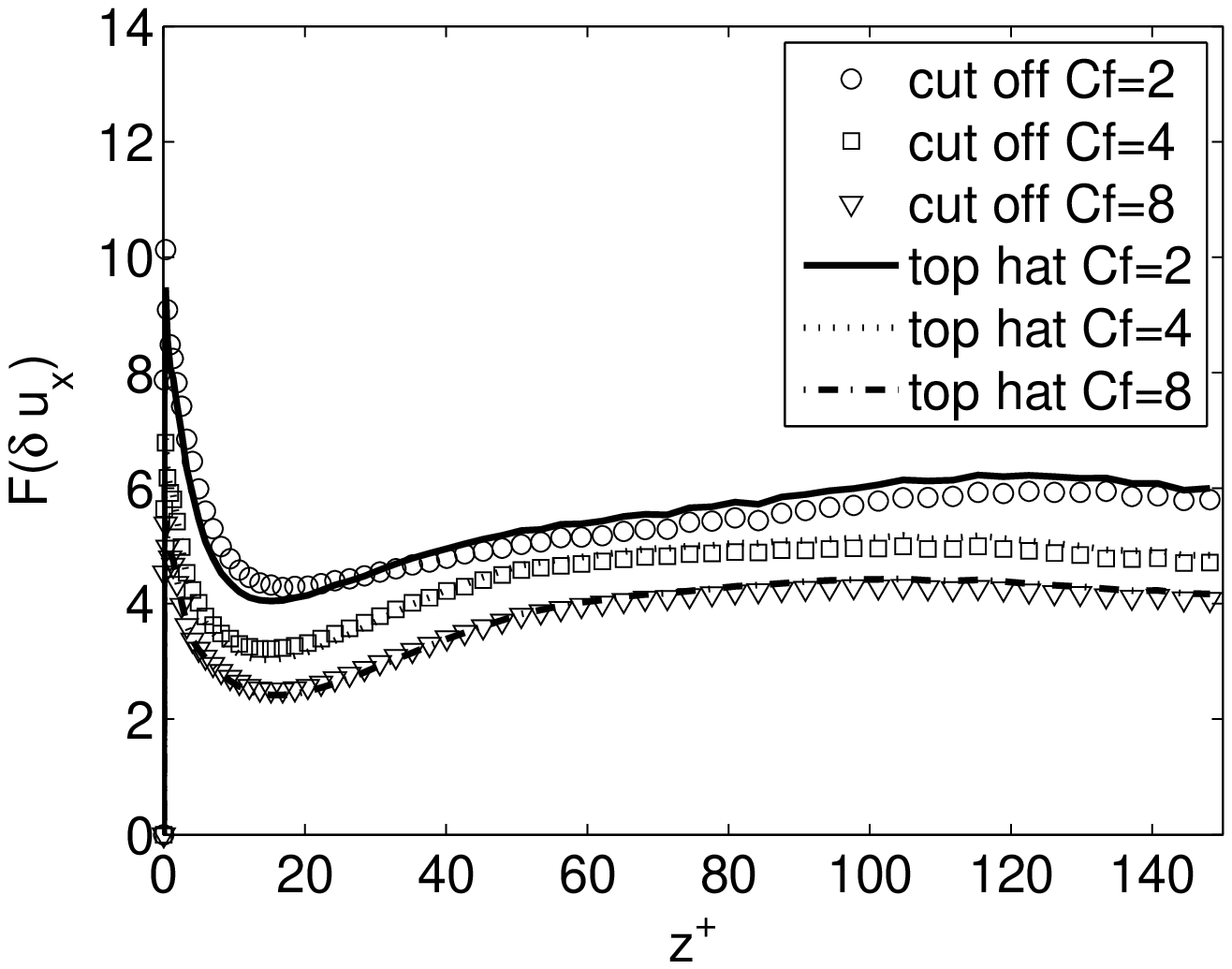}}\\
(a) \hspace{8.0cm} (b)\\
\centerline{\includegraphics[height=6.cm,width=8.cm,angle=0.]{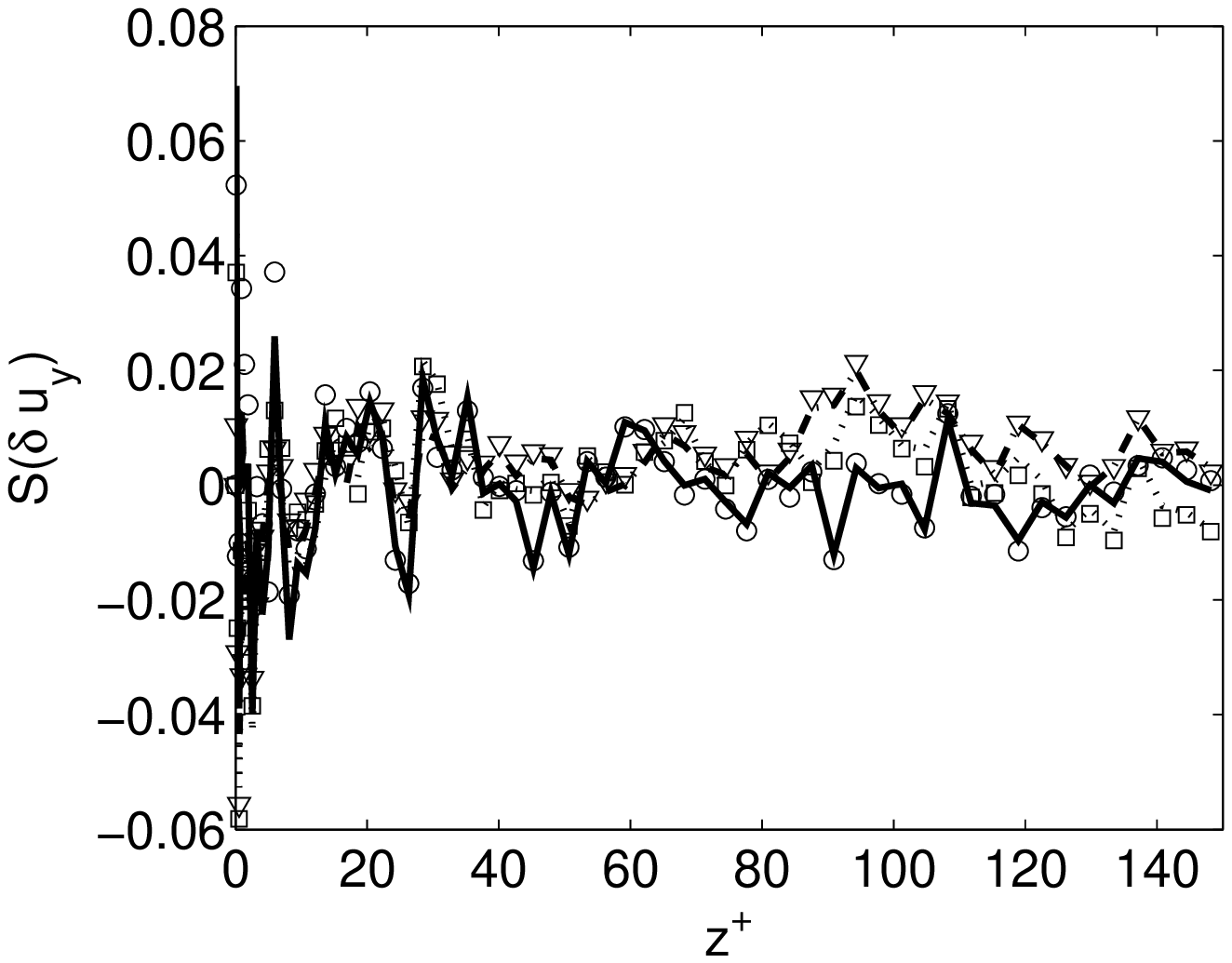}
\includegraphics[height=6.cm,width=8.cm,angle=0.]{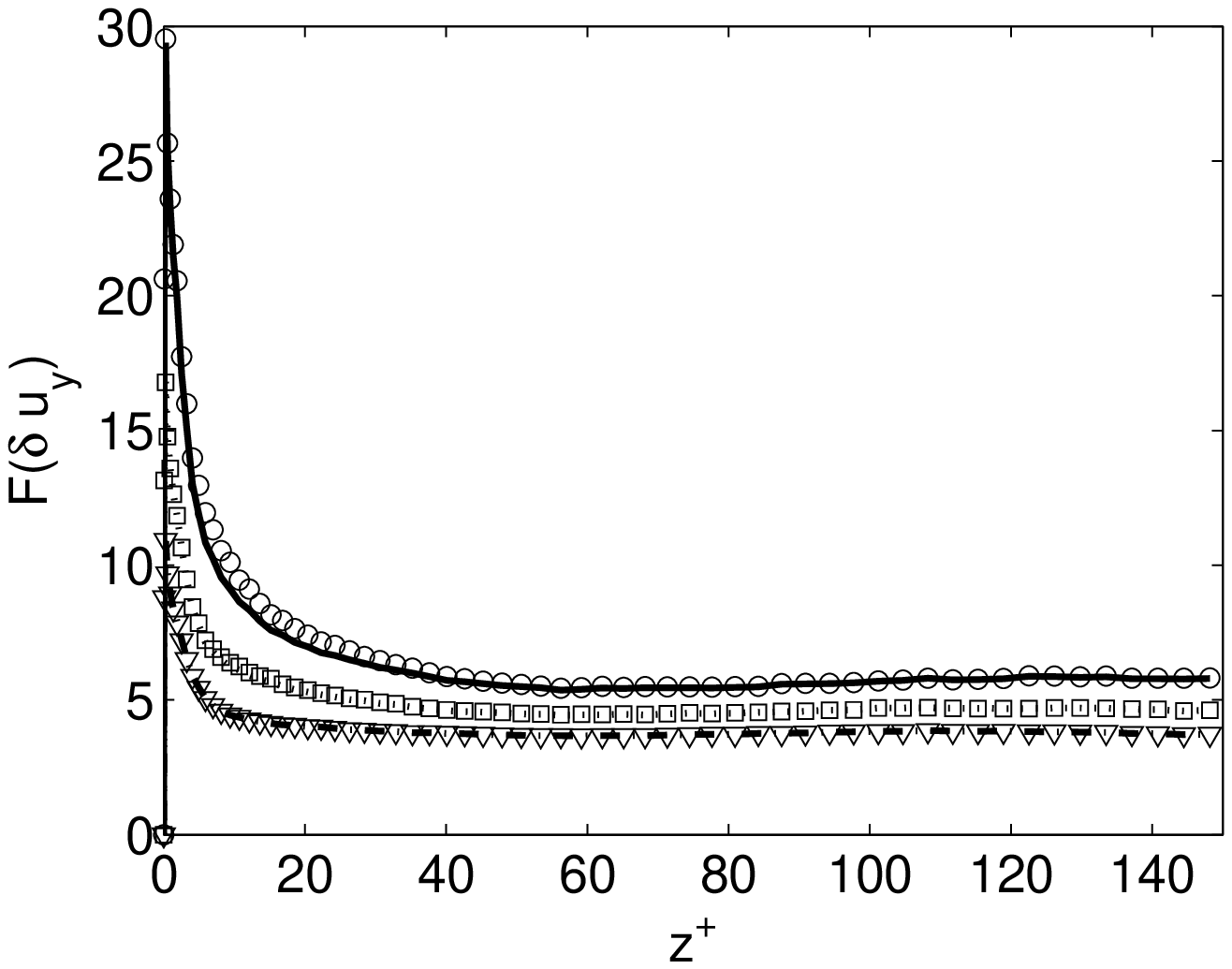}}\\
(c) \hspace{8.0cm} (d)\\
\centerline{\includegraphics[height=6.cm,width=8.cm,angle=0.]{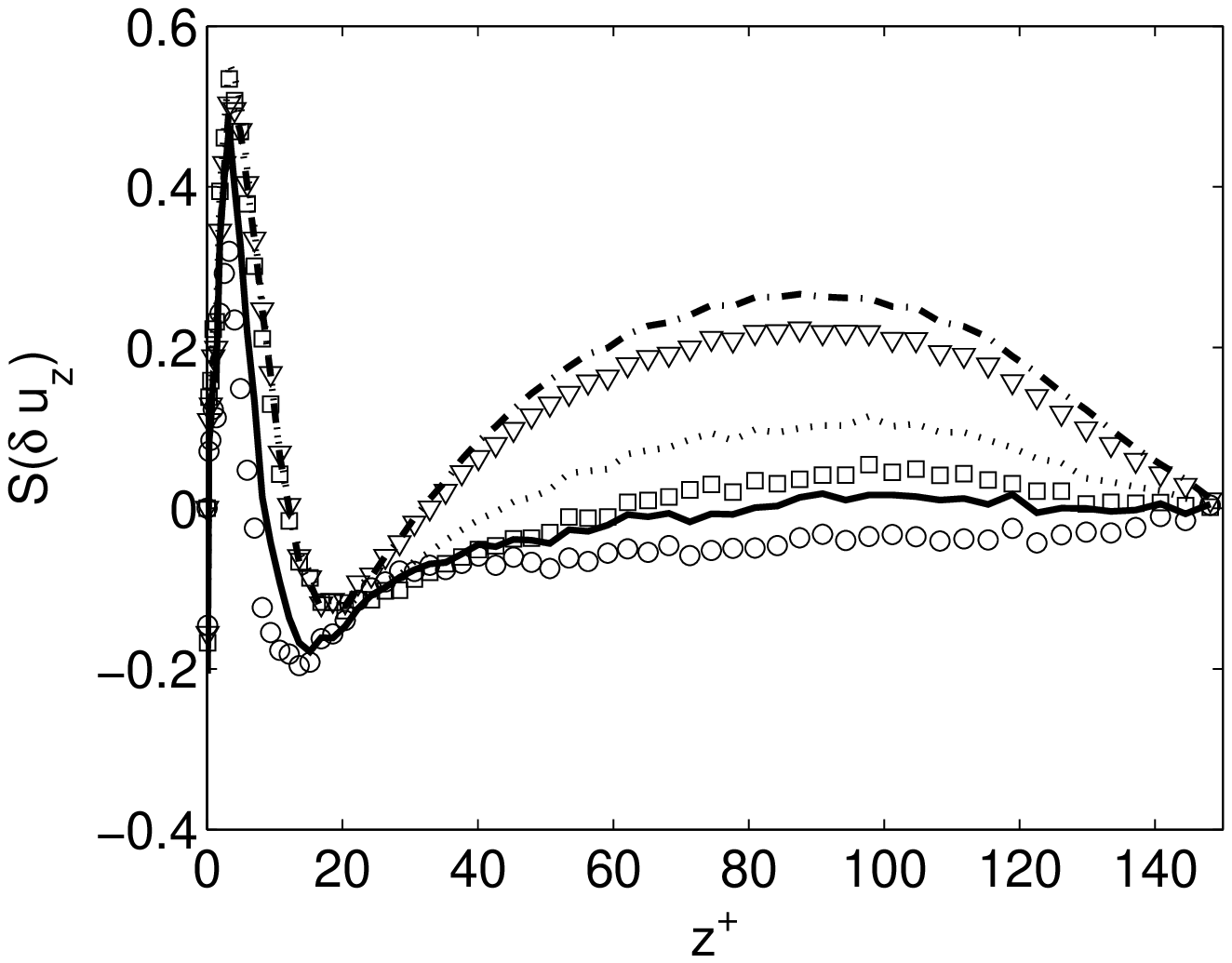}
\includegraphics[height=6.cm,width=8.cm,angle=0.]{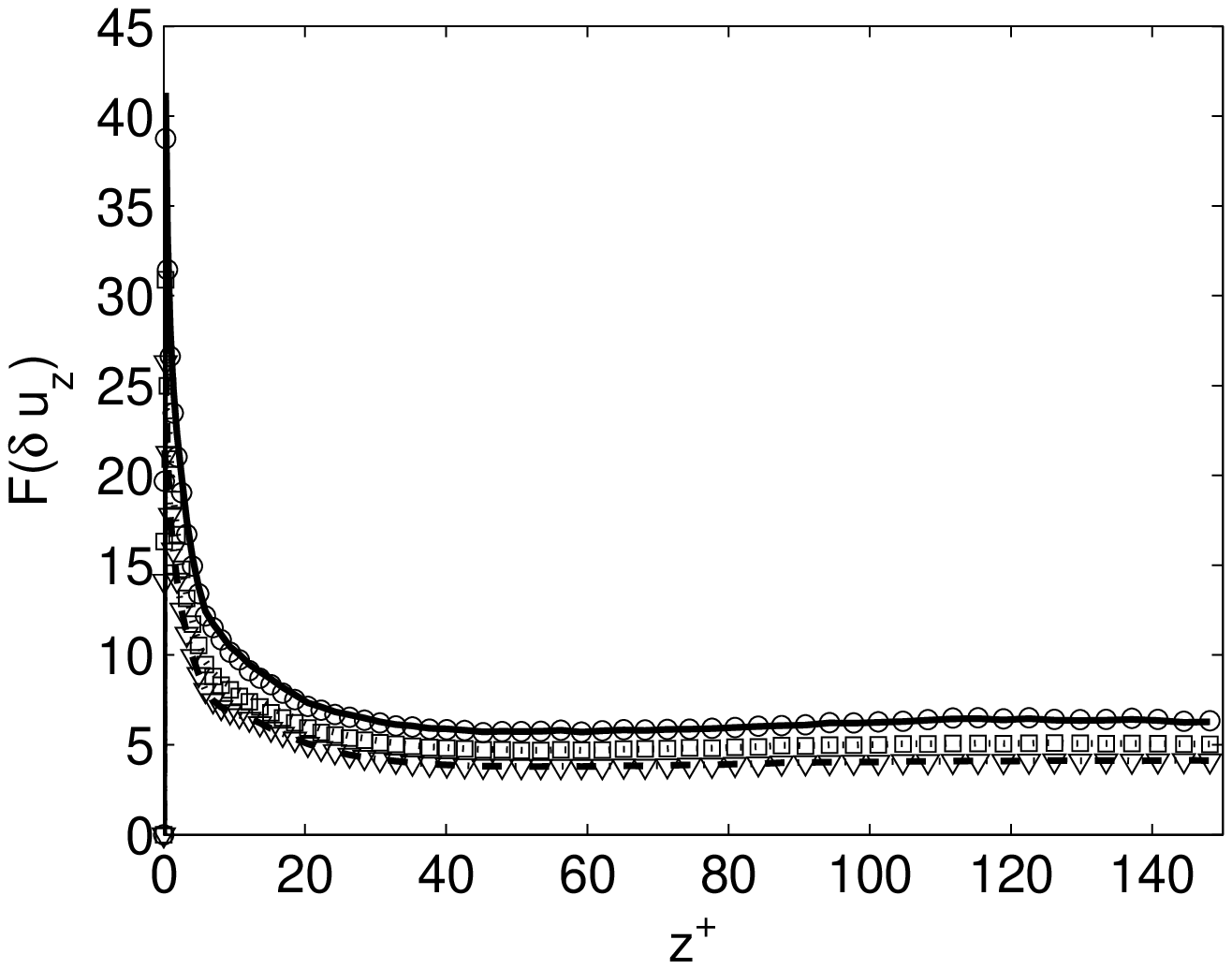}}\\
(e) \hspace{8.0cm} (f)\\
\end{tabular}
\vspace{0.2cm}
\caption{Skewness and flatness of the filtering error $\delta {\bf u}$
in the streamwise (a-b), spanwise (c-d) and wall-normal (e-f) directions
as a function of $z^+$ at varying filter type and width.
Profiles refer to the $St=5$ particles.}
\label{skw_IC_filter}
\end{figure}

\begin{figure}
\begin{tabular}{cc}
\includegraphics[height=6.cm,width=7.cm,angle=0.]{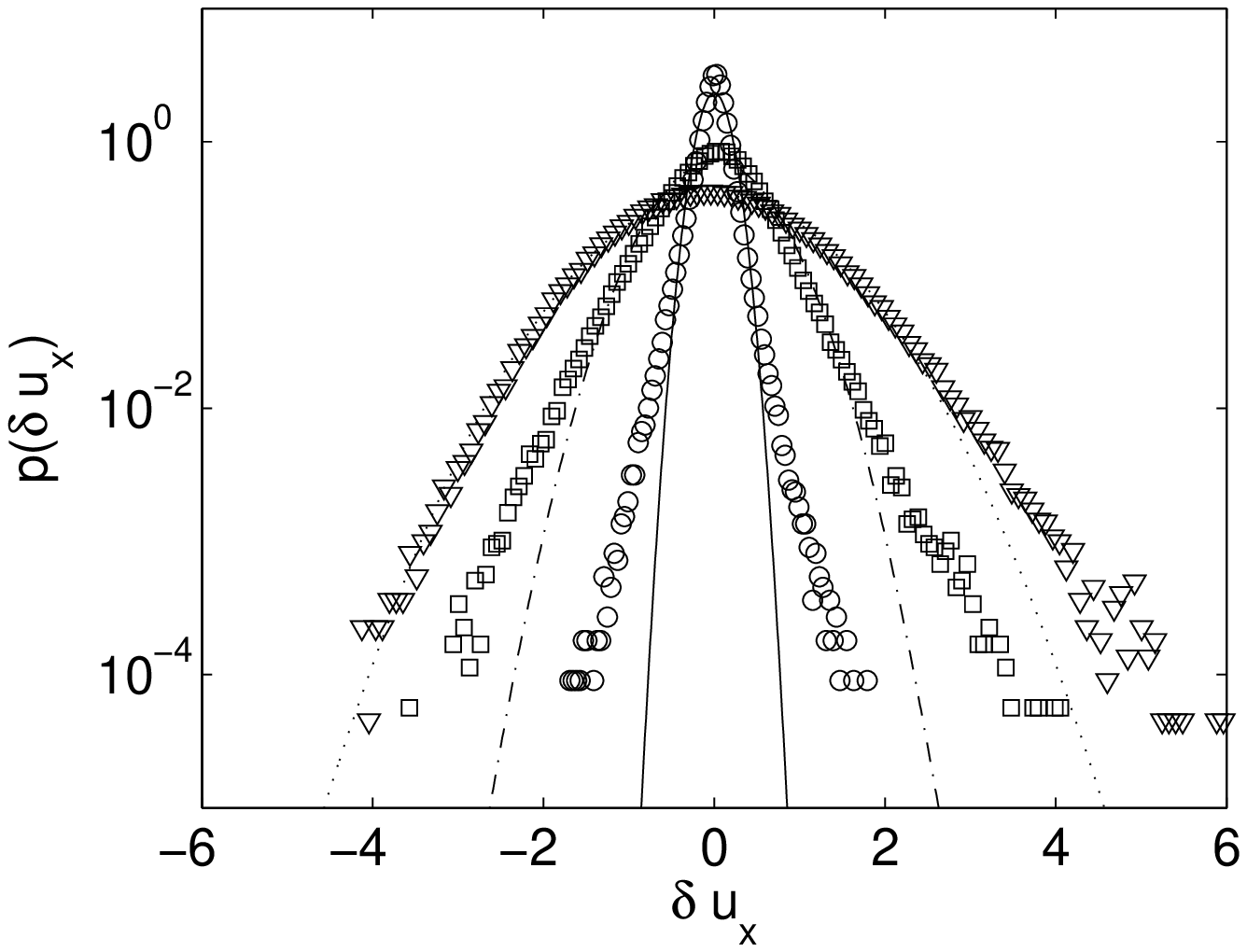} & \includegraphics[height=6.cm,width=7.cm,angle=0.]{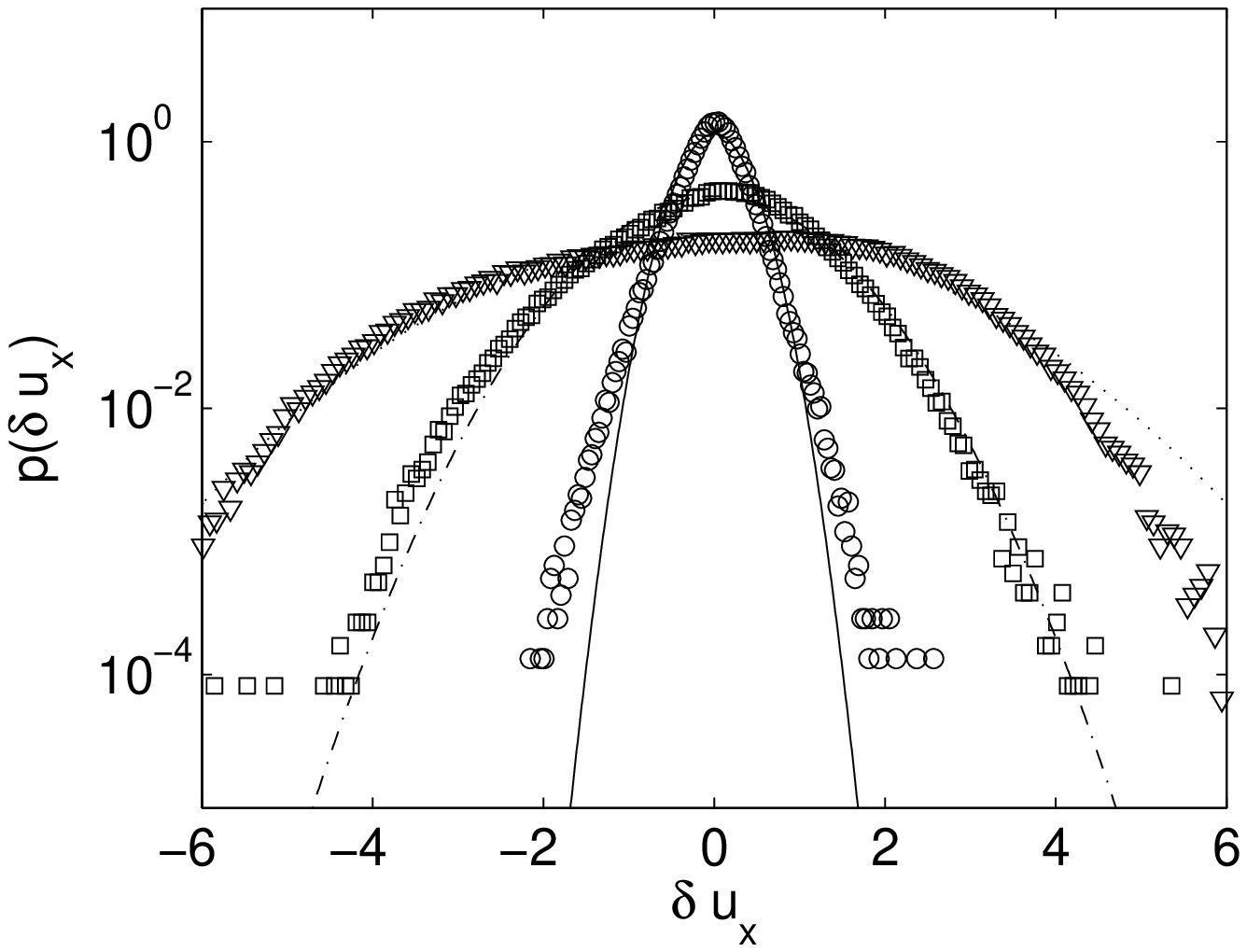}\\
$z^+=4.07$ & $z^+=16.86$\\
\includegraphics[height=6.cm,width=7.cm,angle=0.]{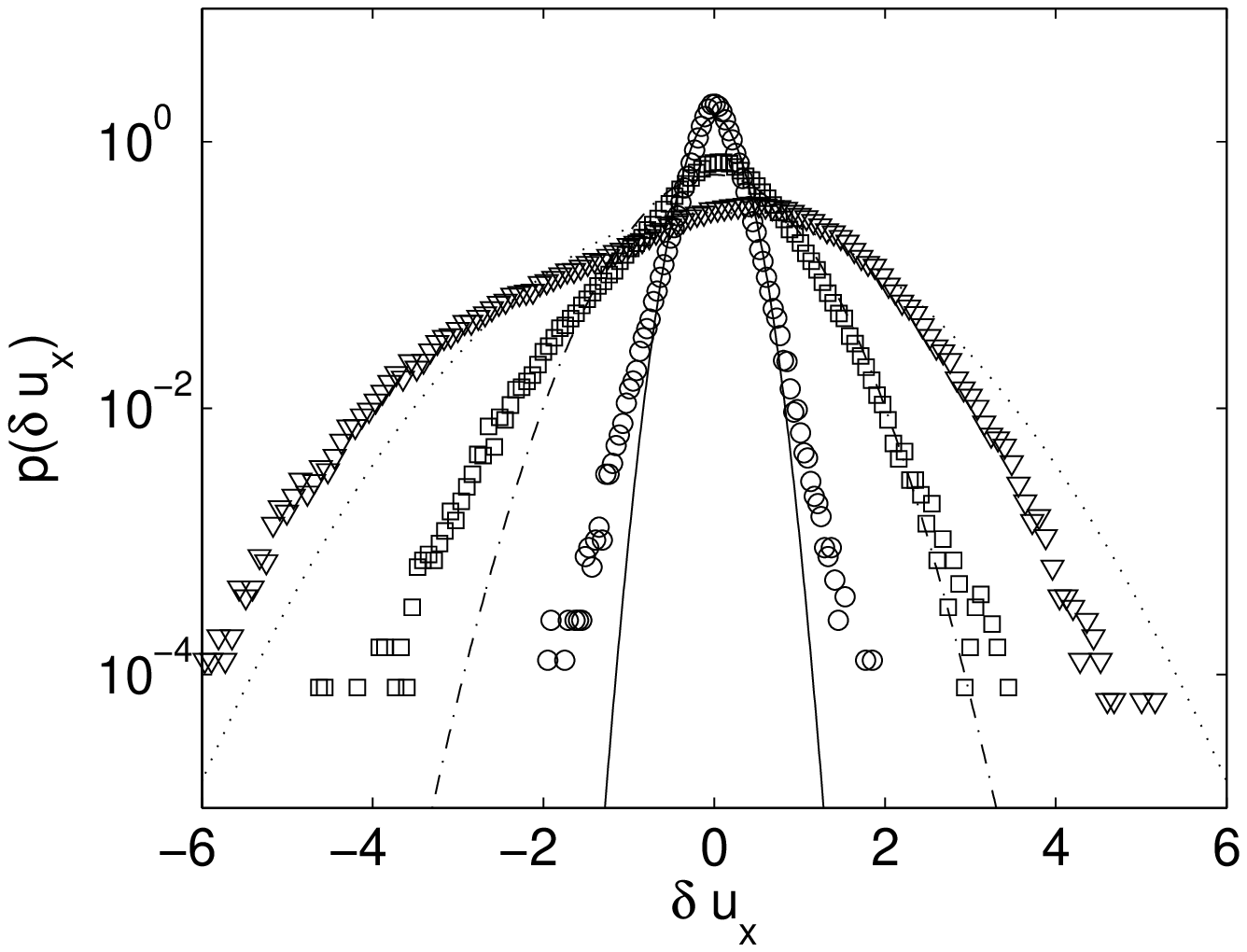} & \includegraphics[height=6.cm,width=7.cm,angle=0.]{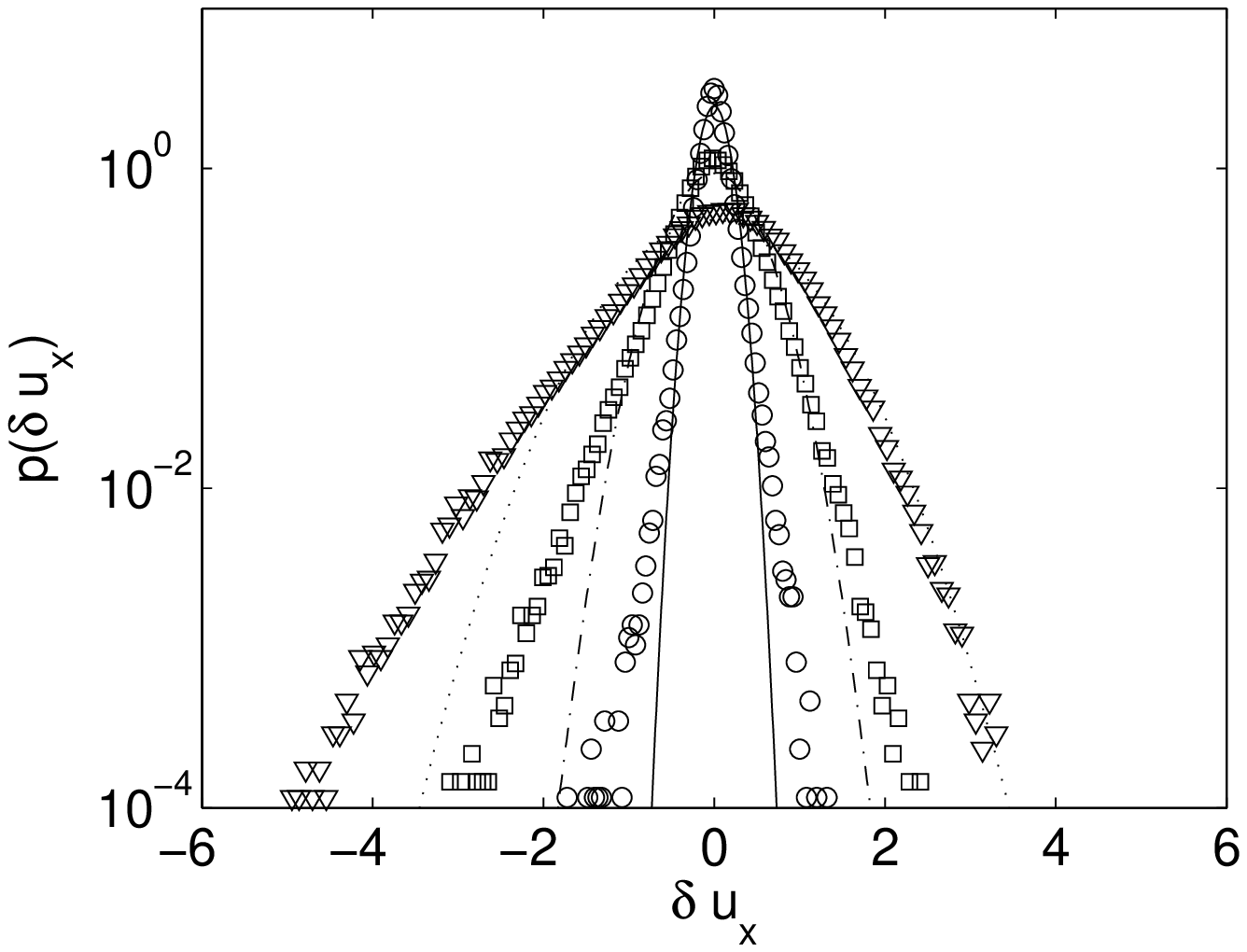}\\
 $z^+=37.64$ & $z^+=65.14$\\
\end{tabular}
\caption{Probability density functions of the streamwise component of the filtering
error for different filter widths. Profiles refer to results obtained for the
$St=5$ particles with cut-off filter.
Open symbols are used for the computed PDFs
($\circ$: CF=2,
$\square$: CF=4,
$\triangledown$: CF=8);
lines for the corresponding Gaussian PDFs
($---$: CF=2, $- \cdot - \cdot -$: CF=4, $\cdot \cdot \cdot \cdot$: CF=8).
}
\label{pdf_x_co}
\end{figure}

\begin{figure}
\begin{tabular}{cc}
\includegraphics[height=6.cm,width=7.cm,angle=0.]{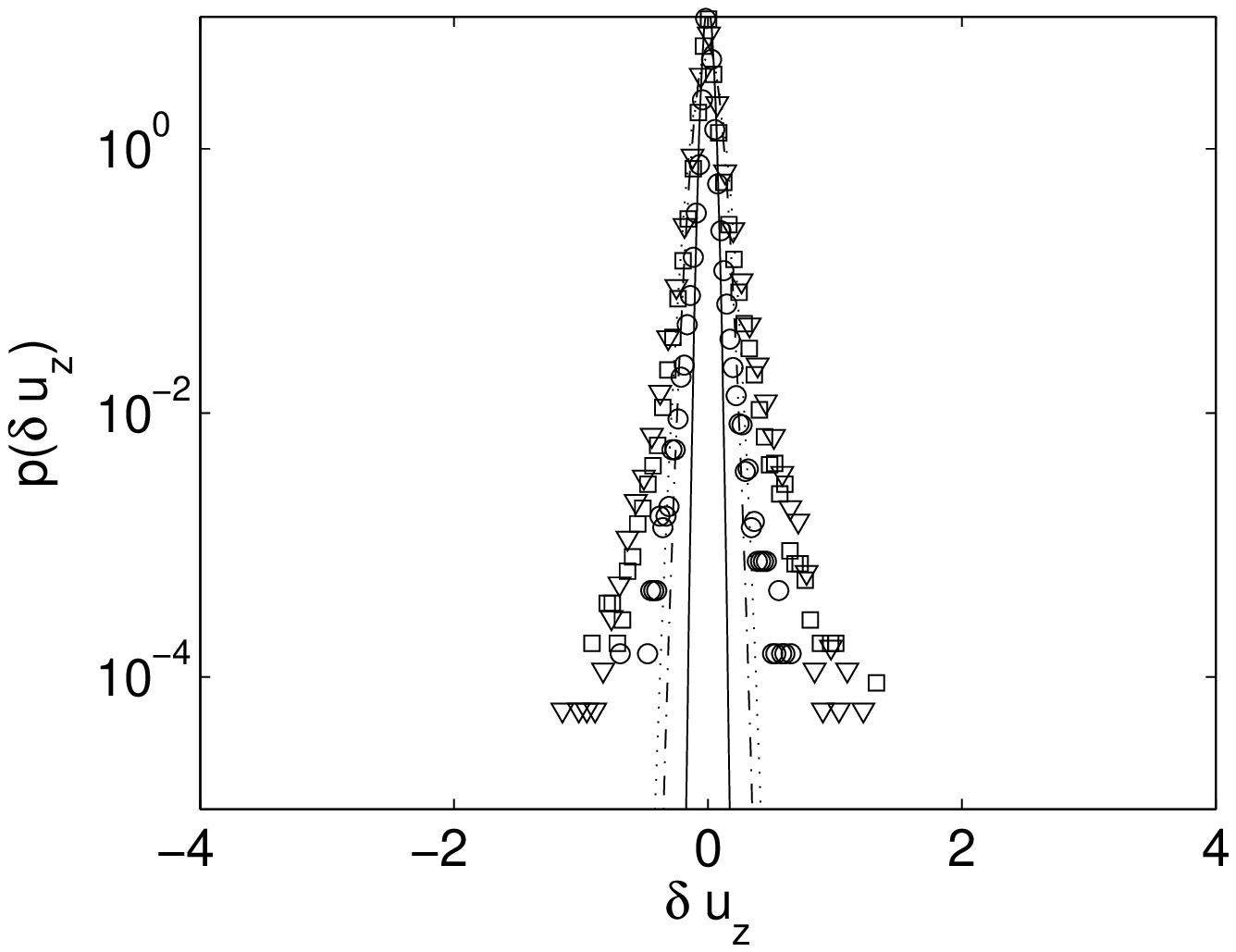} & \includegraphics[height=6.cm,width=7.cm,angle=0.]{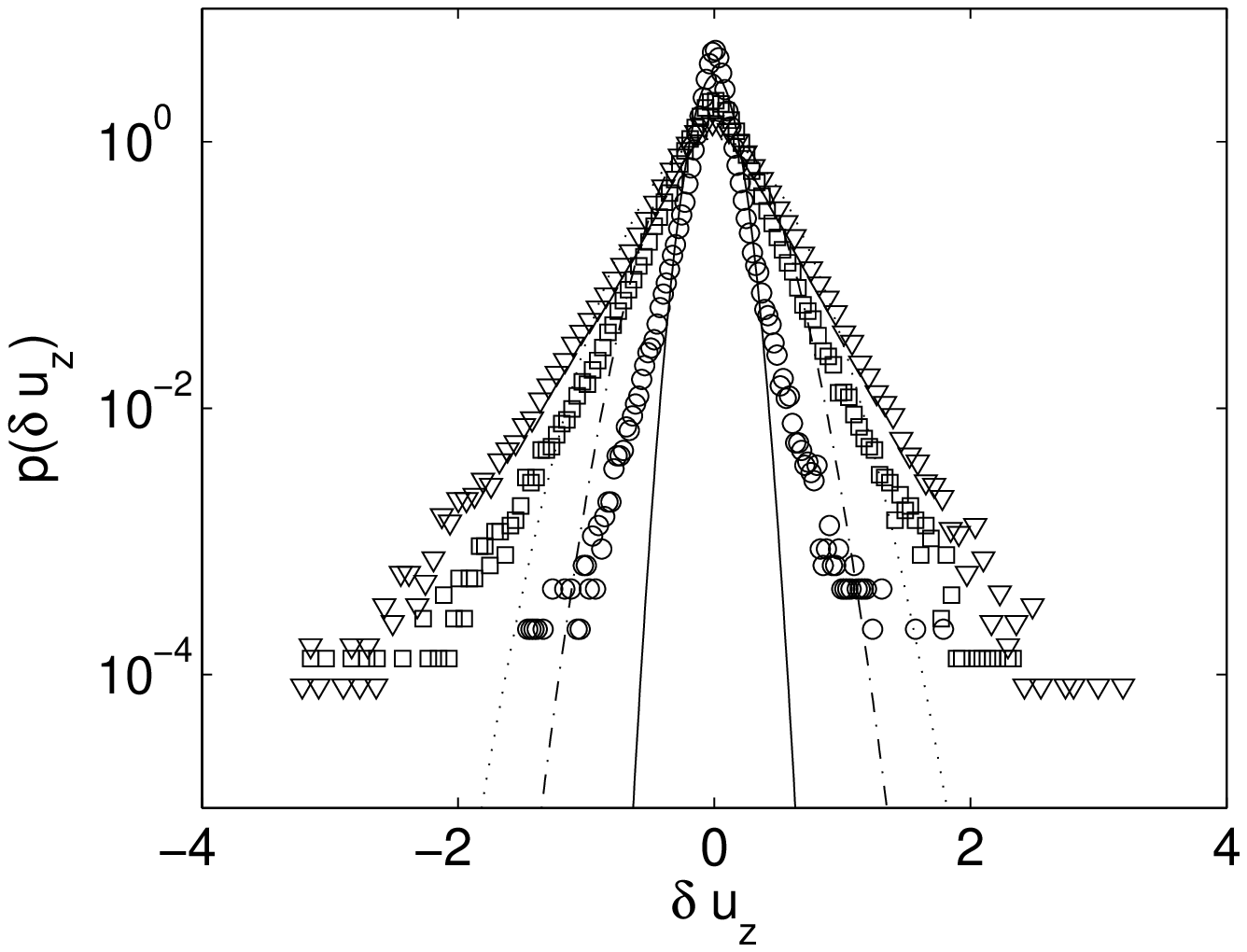}\\
$z^+=4.07$ & $z^+=16.86$\\
\includegraphics[height=6.cm,width=7.cm,angle=0.]{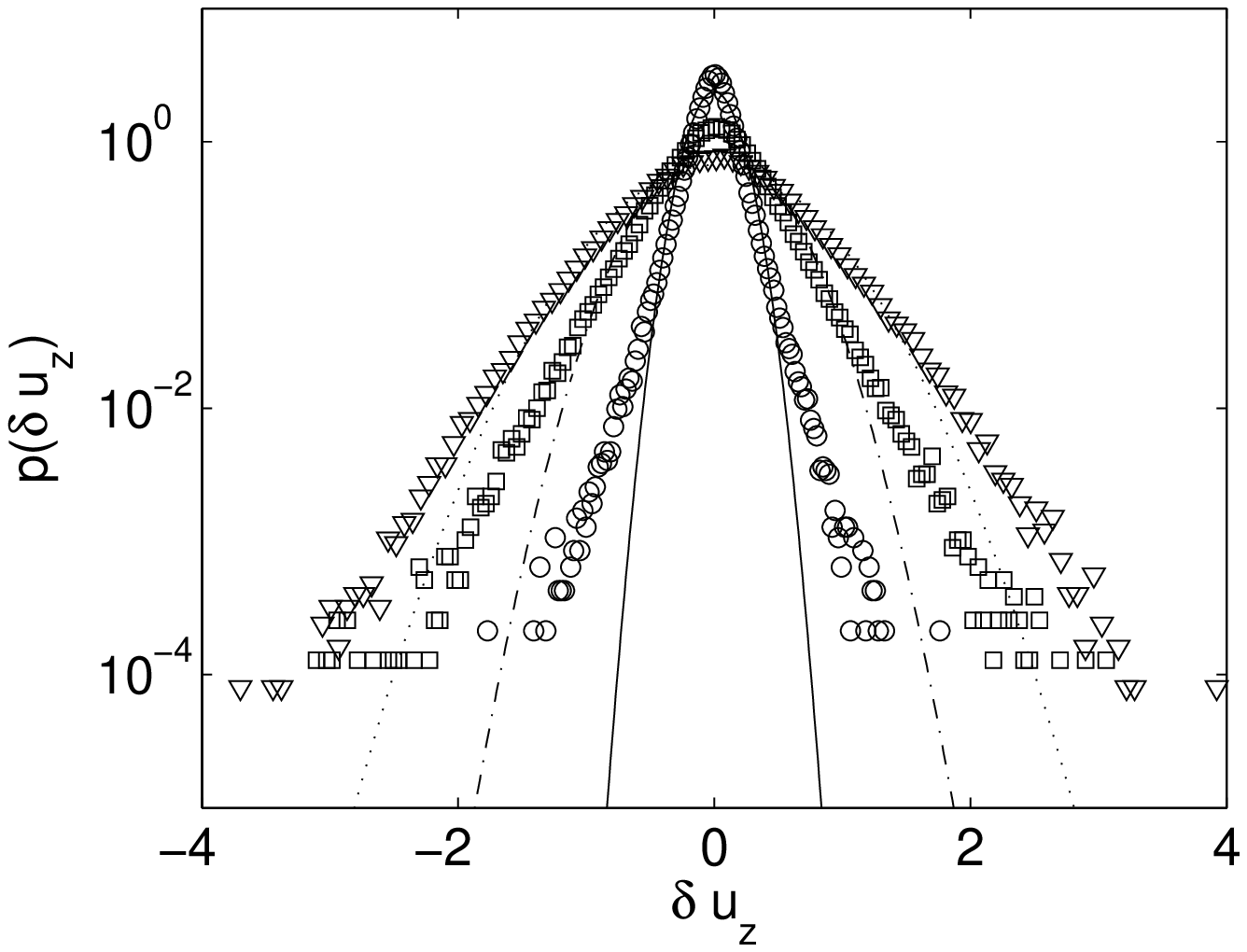} & \includegraphics[height=6.cm,width=7.cm,angle=0.]{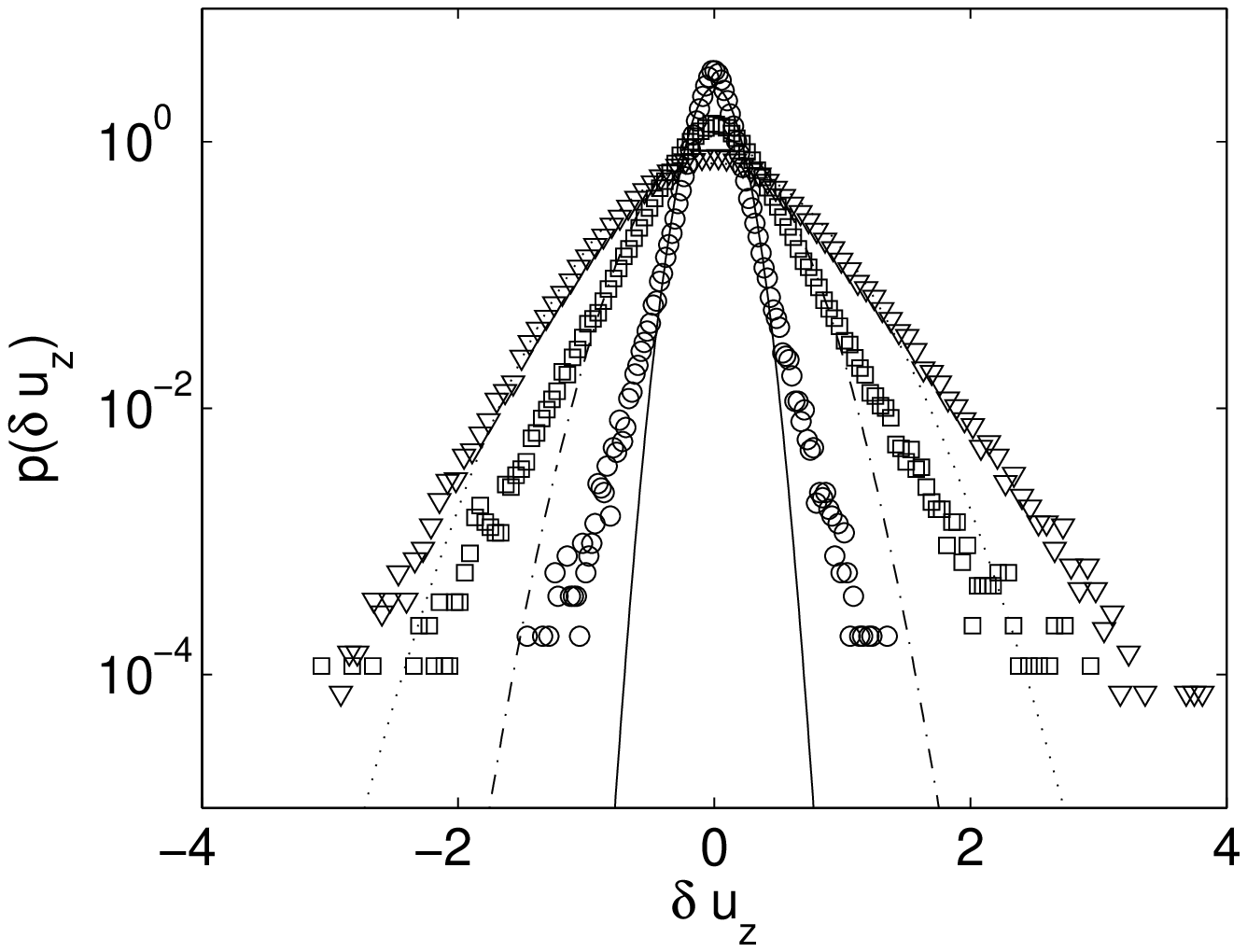}\\
 $z^+=37.64$ & $z^+=65.14$\\
\end{tabular}
\caption{Probability density functions of the wall-normal component of the filtering
error for different filter widths. Profiles refer to results obtained for the
$St=5$ particles with cut-off filter.
Open symbols are used for the computed PDFs
($\circ$: CF=2,
$\square$: CF=4,
$\triangledown$: CF=8);
lines for the corresponding Gaussian PDFs
($---$: CF=2, $- \cdot - \cdot -$: CF=4, $\cdot \cdot \cdot \cdot$: CF=8).
}
\label{pdf_z_co}
\end{figure}

\begin{figure}
\begin{tabular}{cc}
\includegraphics[height=6.cm,width=7.cm,angle=0.]{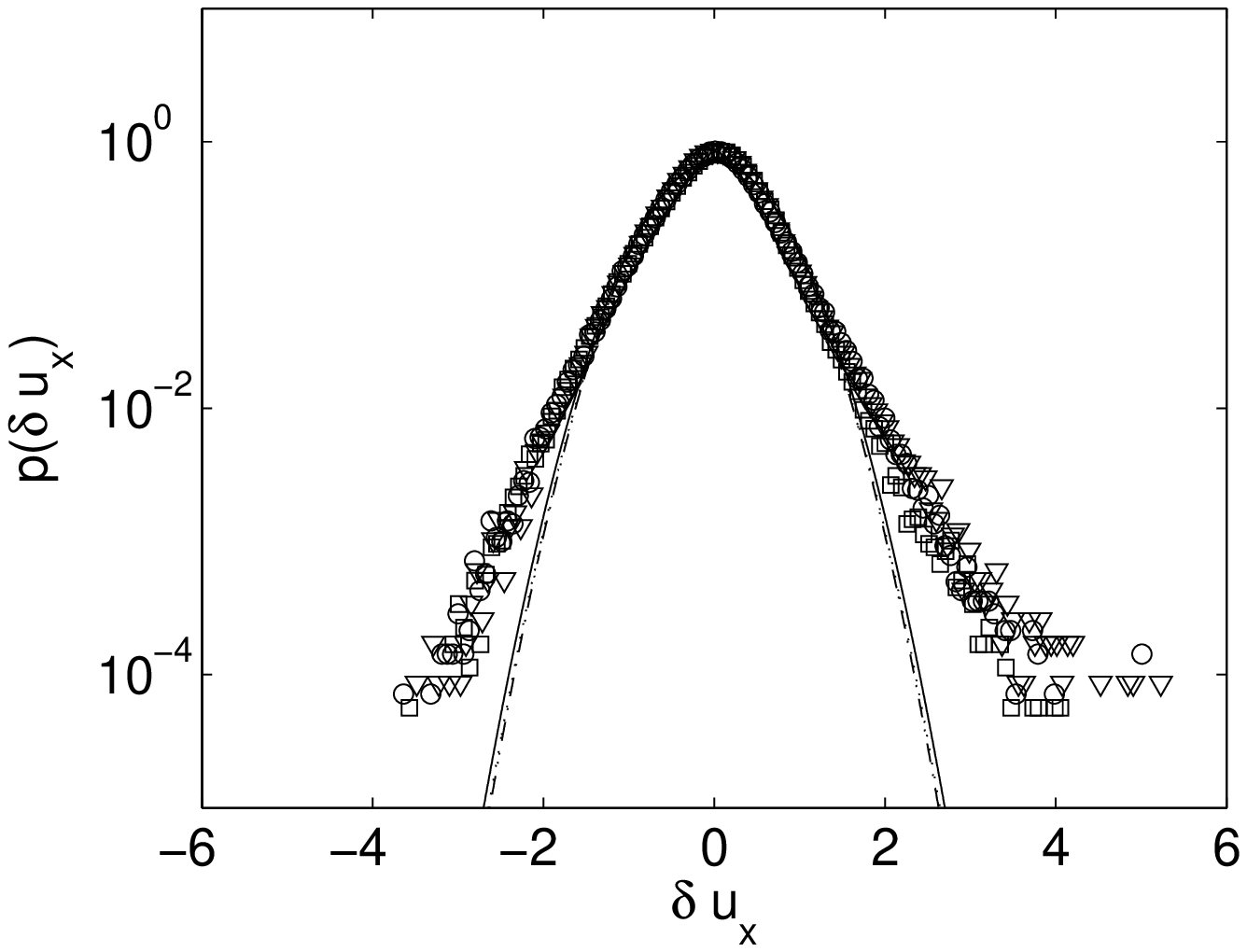} & \includegraphics[height=6.cm,width=7.cm,angle=0.]{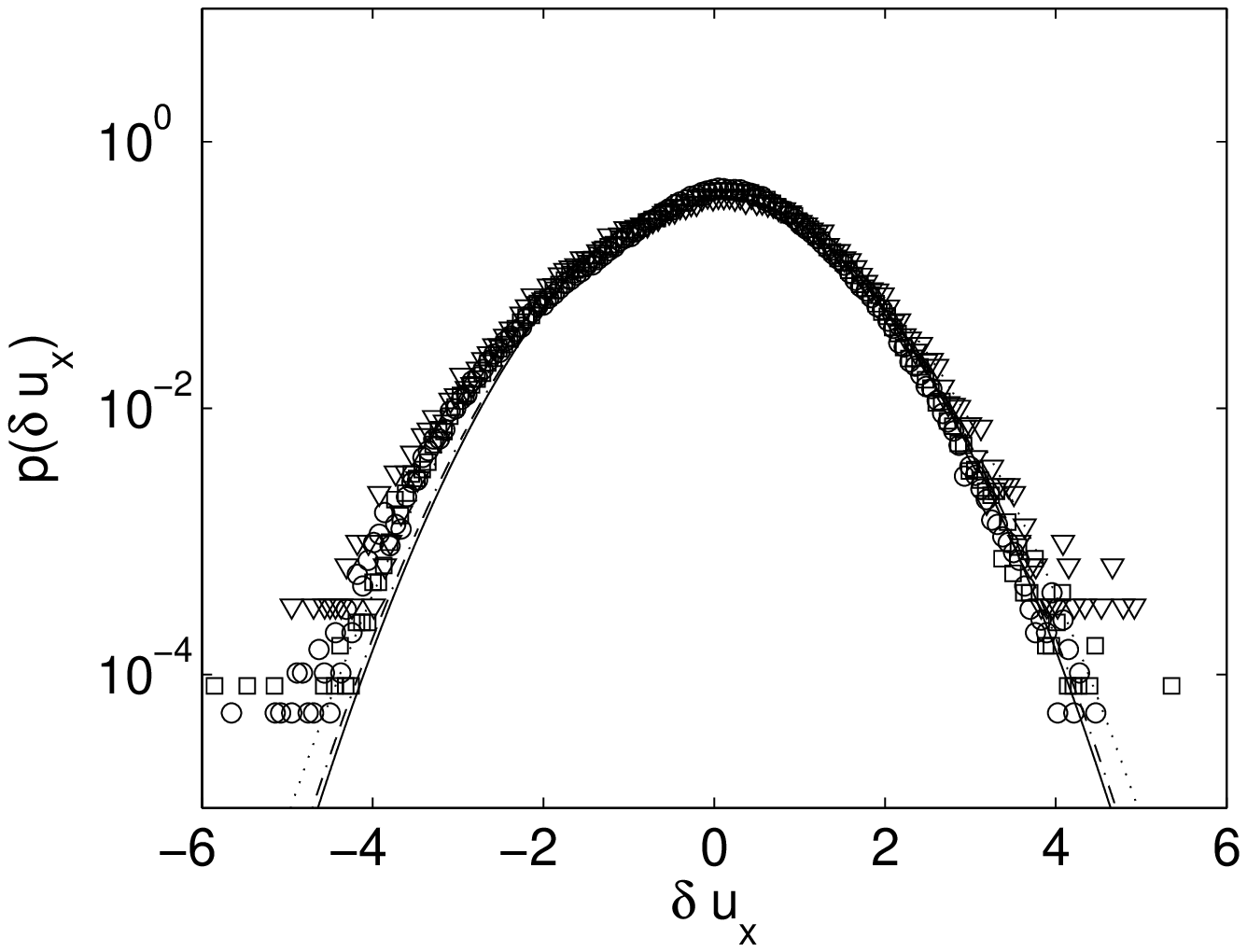}\\
$z^+=4.07$ & $z^+=16.86$\\
\includegraphics[height=6.cm,width=7.cm,angle=0.]{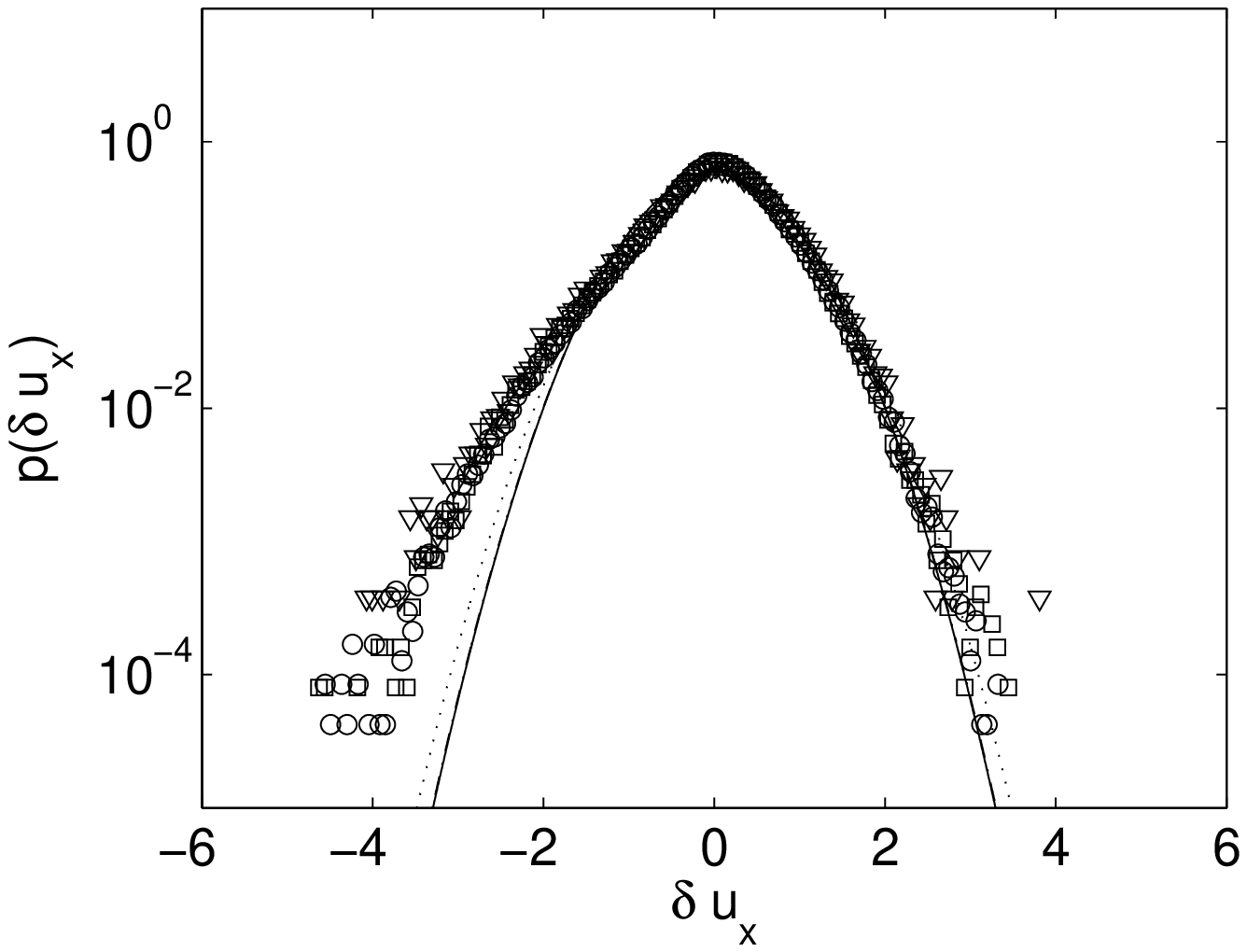} & \includegraphics[height=6.cm,width=7.cm,angle=0.]{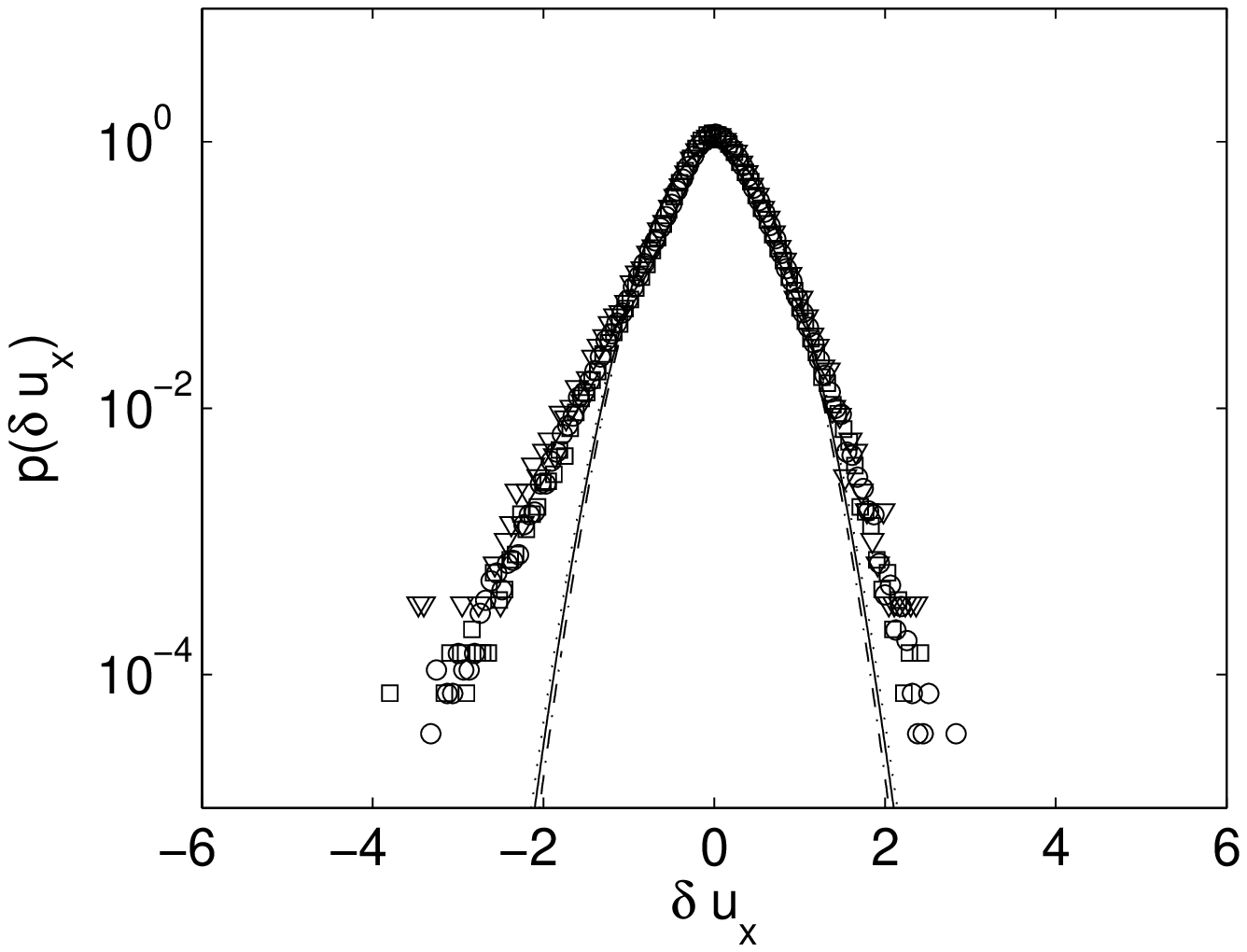}\\
 $z^+=37.64$ & $z^+=65.14$\\
\end{tabular}
\caption{Probability density functions of the streamwise component of the filtering
error for different particle inertia. Profiles refer to results obtained using
cut-off filter with CF=4.
Open symbols are used for the computed PDFs
($\circ$: $St=1$,
$\square$: $St=5$,
$\triangledown$: $St=25$);
lines for the corresponding Gaussian PDFs
($---$: $St=1$, $- \cdot - \cdot -$: $St=5$, $\cdot \cdot \cdot \cdot$: $St=25$).
}
\label{pdf_x_CF4}
\end{figure}

\begin{figure}
\begin{tabular}{cc}
\includegraphics[height=6.cm,width=7.cm,angle=0.]{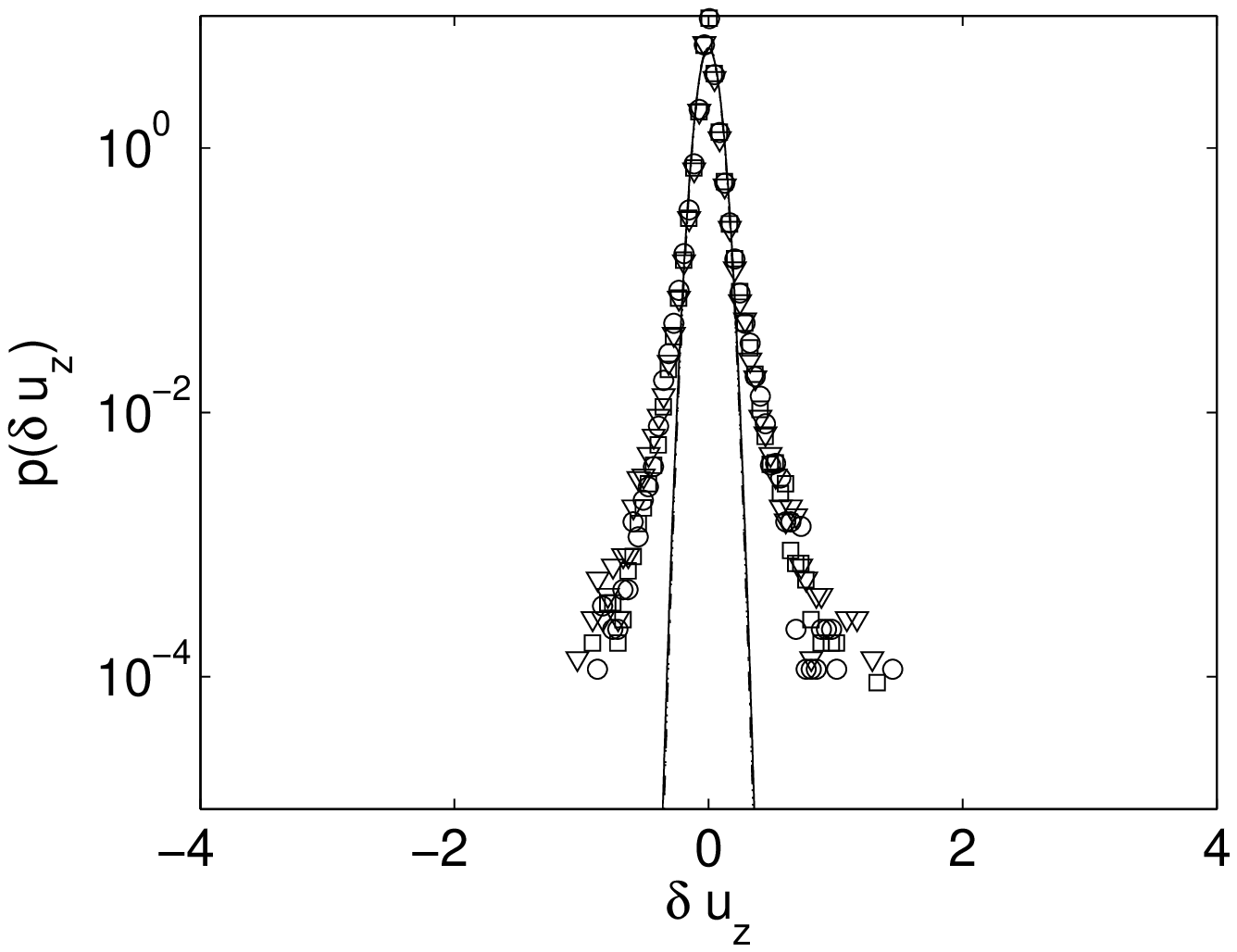} & \includegraphics[height=6.cm,width=7.cm,angle=0.]{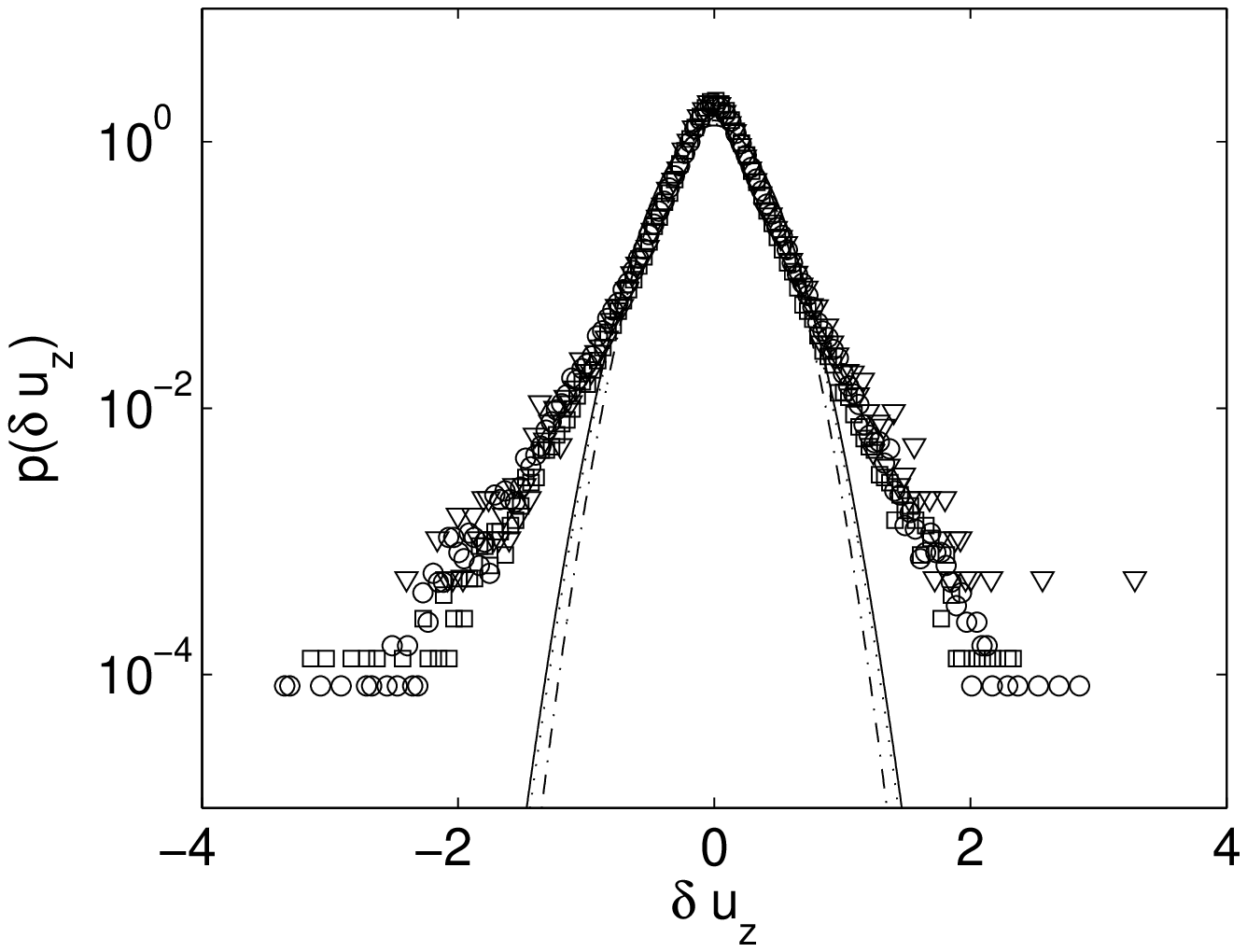}\\
$z^+=4.07$ & $z^+=16.86$\\
\includegraphics[height=6.cm,width=7.cm,angle=0.]{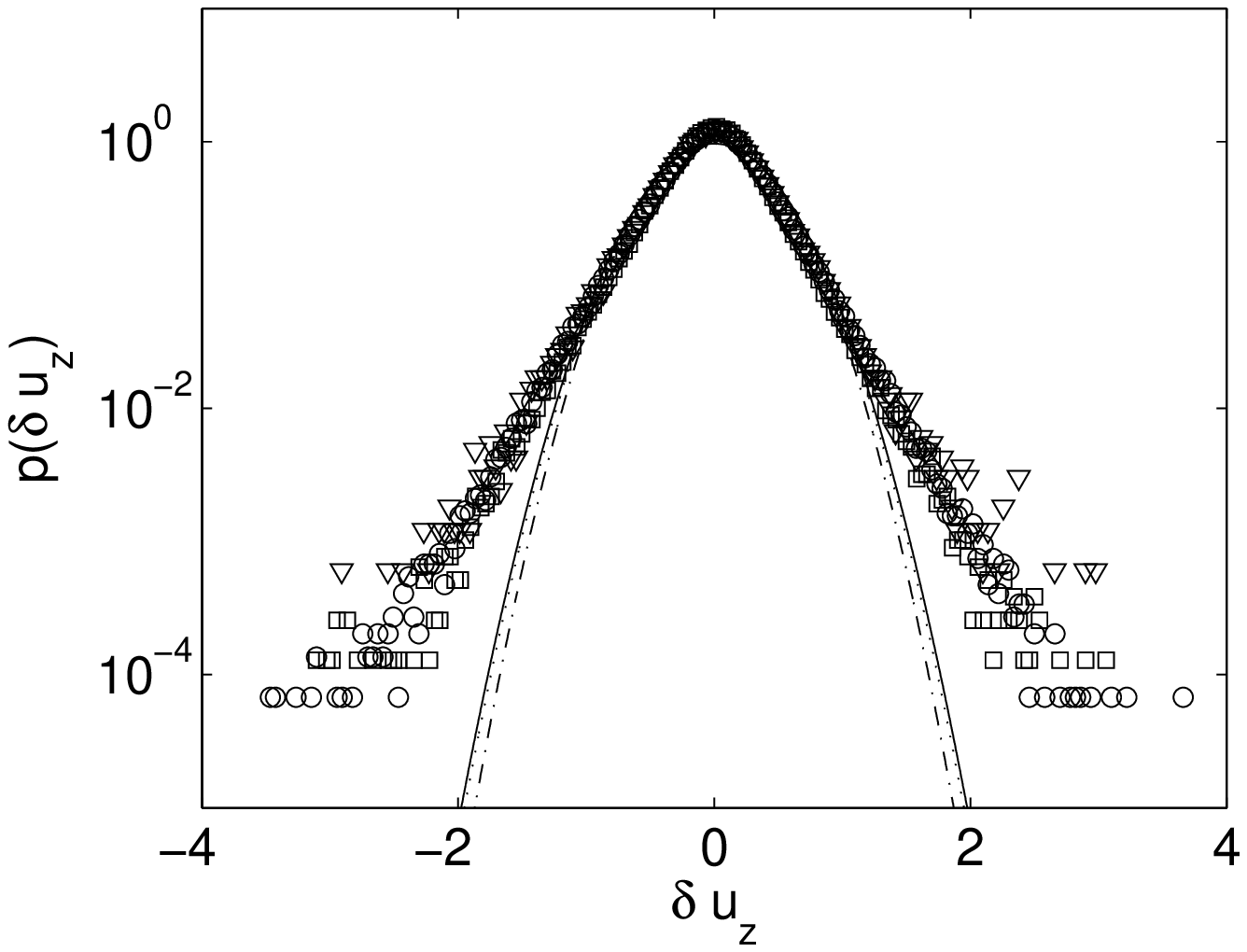} & \includegraphics[height=6.cm,width=7.cm,angle=0.]{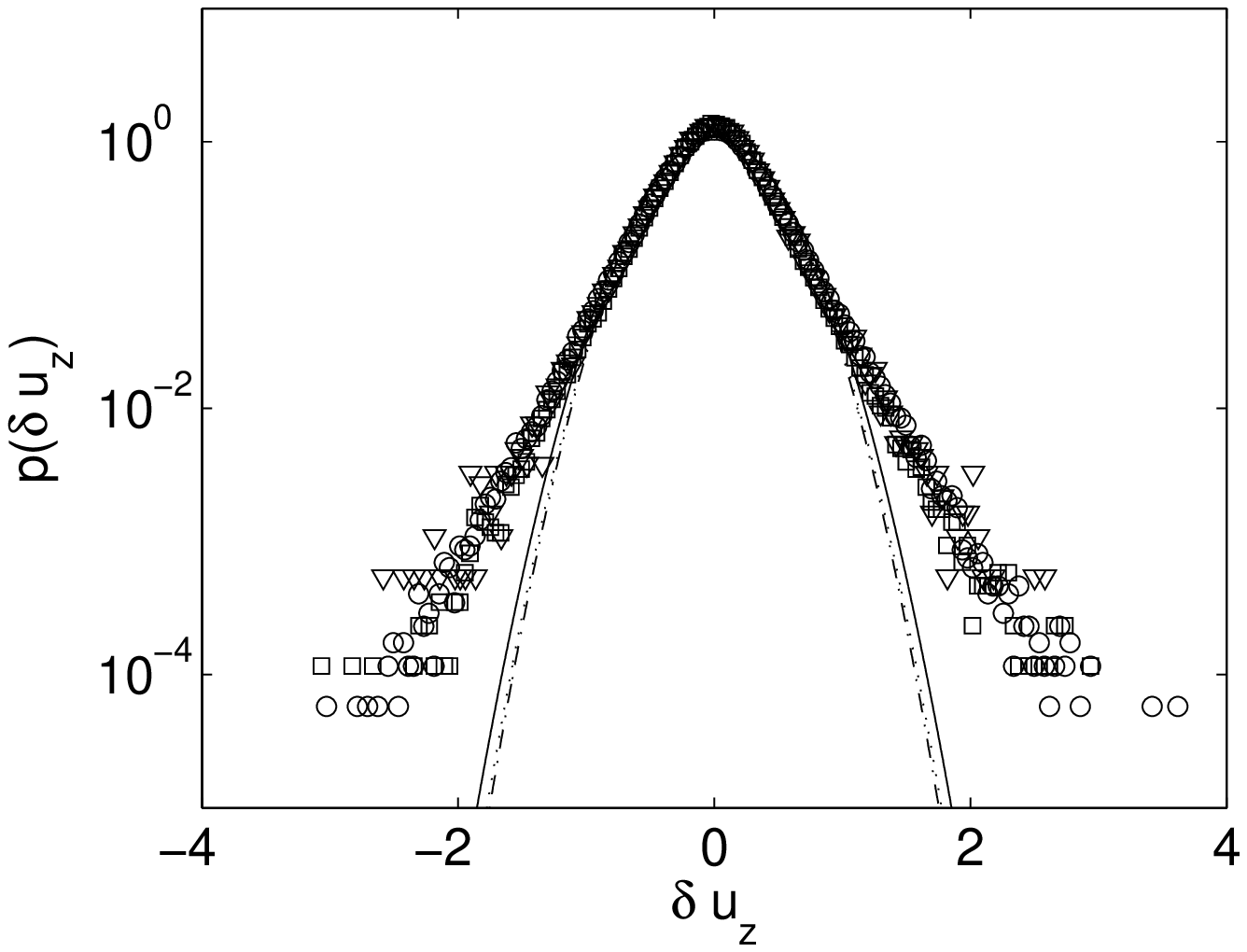}\\
 $z^+=37.64$ & $z^+=65.14$\\
\end{tabular}
\caption{Probability density functions of the wall-normal component of the filtering
error for different particle inertia. Profiles refer to results obtained using
cut-off filter with CF=4.
Open symbols are used for the computed PDFs
($\circ$: $St=1$,
$\square$: $St=5$,
$\triangledown$: $St=25$);
lines for the corresponding Gaussian PDFs
($---$: $St=1$, $- \cdot - \cdot -$: $St=5$, $\cdot \cdot \cdot \cdot$: $St=25$).
}
\label{pdf_z_CF4}
\end{figure}

\clearpage
\newpage
\begin{center}
\textbf{List of Figures}
\end{center}
\noindent Figure \ref{mean_IC_inertia}:
Mean values of the filtering error $\delta {\bf u}$
in the streamwise (a), spanwise (b) and wall-normal (c) directions
as a function of $z^+$ at varying particle inertia.
Profiles refer to the cut-off filter with CF=4.
\\
Figure \ref{fig:streaks}:
Fluid streaks in the near-wall region. Low-speed (resp. high-speed) streaks
are rendered using colored contours of negative (resp. positive) streamwise fluctuating
velocity $u_x'$ on a horizontal plane at $z^+ = 10$ from the wall.
Panel (a) shows the streaky structures obtained from DNS; the
instantaneous distribution of the $St=25$ particles comprised
between $z^+ = 10$ and the wall is also shown to highlight
preferential distribution into the low-speed streaks.
Panels (b) and (c) show the streaky structures obtained from a-priori
DNS with coarsening-factor 4 and 8, respectively.
\\
Figure \ref{fluidvel_part}:
Effects of filtering on the mean wall-normal fluid velocity seen by the
particles at varying particle inertia. Profiles refer to the cut-off filter with CF=4.
\\
Figure \ref{rms_IC_inertia}:
Rms of the filtering error $\delta {\bf u}$ in the streamwise (a),
spanwise (b) and wall-normal (c) directions as a function of $z^+$ at
varying particle inertia. Profiles
refer to the cut-off filter with CF=4.
\\
Figure \ref{skew_IC_inertia}:
Skewness and flatness of the filtering error $\delta {\bf u}$
in the streamwise (a-b), spanwise (c-d) and wall-normal (e-f) directions
as a function of $z^+$ at
varying particle inertia. Profiles refer to the cut-off filter with CF=4.
\\
Figure \ref{mean_IC_filter}:
Mean values of the filtering error $\delta {\bf u}$
in the streamwise (a), and wall-normal (b) directions
as a function of $z^+$ at varying filter type and width.
Profiles refer to the $St=5$ particles.
\\
Figure \ref{rms_IC_filter}:
Rms of the filtering error $\delta {\bf u}$
in the streamwise (a), spanwise (b) and normal (c) directions
as a function of $z^+$ at varying filter type and width.
Profiles refer to the $St=5$ particles.
\\
Figure \ref{skw_IC_filter}:
Skewness and flatness of the filtering error $\delta {\bf u}$
in the streamwise (a-b), spanwise (c-d) and wall-normal (e-f) directions
as a function of $z^+$ at varying filter type and width.
Profiles refer to the $St=5$ particles.
\\
Figure \ref{pdf_x_co}:
Probability density functions of the streamwise component of the filtering
error for different filter widths. Profiles refer to results obtained for the
$St=5$ particles with cut-off filter.
Open symbols are used for the computed PDFs
($\circ$: CF=2,
$\square$: CF=4,
$\triangledown$: CF=8);
lines for the corresponding Gaussian PDFs
($---$: CF=2, $- \cdot - \cdot -$: CF=4, $\cdot \cdot \cdot \cdot$: CF=8).
\\
Figure \ref{pdf_z_co}:
Probability density functions of the wall-normal component of the filtering
error for different filter widths. Profiles refer to results obtained for the
$St=5$ particles with cut-off filter.
Open symbols are used for the computed PDFs
($\circ$: CF=2,
$\square$: CF=4,
$\triangledown$: CF=8);
lines for the corresponding Gaussian PDFs
($---$: CF=2, $- \cdot - \cdot -$: CF=4, $\cdot \cdot \cdot \cdot$: CF=8).
\\
Figure \ref{pdf_x_CF4}:
Probability density functions of the streamwise component of the filtering
error for different particle inertia. Profiles refer to results obtained using
cut-off filter with CF=4.
Open symbols are used for the computed PDFs
($\circ$: $St=1$,
$\square$: $St=5$,
$\triangledown$: $St=25$);
lines for the corresponding Gaussian PDFs
($---$: $St=1$, $- \cdot - \cdot -$: $St=5$, $\cdot \cdot \cdot \cdot$: $St=25$).
\\
Figure \ref{pdf_z_CF4}:
Probability density functions of the wall-normal component of the filtering
error for different particle inertia. Profiles refer to results obtained using
cut-off filter with CF=4.
Open symbols are used for the computed PDFs
($\circ$: $St=1$,
$\square$: $St=5$,
$\triangledown$: $St=25$);
lines for the corresponding Gaussian PDFs
($---$: $St=1$, $- \cdot - \cdot -$: $St=5$, $\cdot \cdot \cdot \cdot$: $St=25$).
\\
\newpage


\clearpage
\newpage

\newpage
\newpage
\newpage
%
%
\newpage
\newpage
\newpage
%

\newpage


\newpage

\newpage

\newpage
\end{document}